\documentclass[10pt,conference]{IEEEtran}

\usepackage[numbers]{natbib}

\usepackage{enumitem}
\usepackage{comment}
\usepackage{amsmath}
\usepackage{graphicx}
\usepackage{listings}
\usepackage{booktabs}
\usepackage{siunitx} 
\usepackage{adjustbox}
\usepackage{xurl}
\usepackage[skins]{tcolorbox}
\usepackage{subfigure}
\usepackage{fontawesome}
\usepackage{multirow}
\usepackage{wrapfig}
\usepackage{ctable}
\usepackage{amssymb}
\usepackage[linesnumbered,ruled,vlined]{algorithm2e}

\usepackage{array}
\newcommand{\PreserveBackslash}[1]{\let\temp=\\#1\let\\=\temp}
\newcolumntype{C}[1]{>{\PreserveBackslash\centering}p{#1}}
\newcolumntype{R}[1]{>{\PreserveBackslash\raggedleft}p{#1}}
\newcolumntype{L}[1]{>{\setlength{\baselineskip}{0.7\baselineskip}\PreserveBackslash\raggedright}p{#1}}

\newcommand{\TableLineExtraSpace}{6pt}

\AtBeginDocument{%
  \providecommand\BibTeX{{%
\normalfont B\kern-0.5em{\scshape i\kern-0.25em b}\kern-0.8em\TeX}}}

\renewcommand\paragraph[1]{
\subsubsection*{#1}}

\newcommand{\EMH}[1]{}
\newcommand{\AK}[1]{}
\newcommand{\Code}[1]{
  \begin{small}\texttt{#1}
\end{small}}

\begin{document}

\title{A Taxonomy and Taxonomy-Guided Benchmark Reduction for Spreadsheet Tasks}

\makeatletter
\newcommand{\linebreakand}{%
\end{@IEEEauthorhalign}\hfill\mbox{}\par\mbox{}\hfill
\begin{@IEEEauthorhalign}}
\makeatother

\author{%
  \IEEEauthorblockN{Bogdan Vasilescu}
  \IEEEauthorblockA{\textit{Carnegie Mellon University}\\vasilescu@cmu.edu}
  \and
  \IEEEauthorblockN{Emerson Murphy-Hill}
  \IEEEauthorblockA{\textit{Microsoft}\\emerson.rex@microsoft.com}
  \and
  \IEEEauthorblockN{Anikesh Parashar}
  \IEEEauthorblockA{\textit{Microsoft}\\t-aparashar@microsoft.com}
  \and
  \IEEEauthorblockN{Aayush Kumar}
  \IEEEauthorblockA{\textit{MIT}\\aayushk@mit.edu}
  \linebreakand
  \IEEEauthorblockN{Gustavo Soares}
  \IEEEauthorblockA{\textit{Google} (work done at
  Microsoft)\\gustavoasoares@outlook.com}
  \and
  \IEEEauthorblockN{Mar Ginés Marín}
  \IEEEauthorblockA{\textit{Independent} (work done at
  Microsoft)\\mar\_gines@hotmail.com}
  \and
  \IEEEauthorblockN{Sumit Gulwani}
  \IEEEauthorblockA{\textit{Microsoft}\\sumitg@microsoft.com}
}

\maketitle

\bstctlcite{IEEEexample:BSTcontrol}

\begin{abstract}
  Spreadsheets let millions of end users without programming training do computational work, and teams are increasingly building AI agents to help
  them. Evaluating each candidate agent configuration across many tasks is
  costly. Benchmark reduction can lower this cost by selecting a small subset
  of tasks whose outcomes predict full-benchmark scores, but selecting tasks
  only for predictive accuracy can leave whole categories of spreadsheet work
  untested.
  We contribute a taxonomy of spreadsheet tasks, grounded in a review of
  benchmarks and refined with the Excel Agent product team at Microsoft, and a
  benchmark-reduction method that favors taxonomic diversity among similarly
  predictive choices. On benchmark logs from Microsoft, taxonomy-guided
  selection broadens coverage at little or no cost in predictive accuracy and
  often improves prediction on configurations excluded from selection. Broader
  coverage can therefore support, rather than compromise, predictive accuracy.
  Generalization remains limited across time: subsets selected using little
  recent history predict poorly for configurations much later in time. Beyond
  benchmark reduction, the taxonomy provides a reusable framework for auditing
  gaps in task coverage, as teams at Microsoft have done.
\end{abstract}

\section{Introduction}
\label{sec:introduction}

Major technology
companies
and startups are building AI-powered spreadsheet agents: systems that let
users express intent in natural language and autonomously carry out the
corresponding spreadsheet operations.
Academic research on such agents is growing~\citep{li2023sheetcopilot,
chen2025sheetagent, ma2024spreadsheetbench},
mirroring the broader trend toward coding agents in software engineering and
data work~\citep{jimenez2024swebench, liu2023agentbench}, in this case in an
end-user programming environment.
If effective, these agents can democratize complex data operations in a domain
where errors are notoriously
prevalent~\citep{panko1998what, hermans2012detecting}.

Ensuring the quality of these agents is challenging.
Given a goal such as \emph{``clean this sales export and build a
month-over-month revenue summary,''} an agent plans and carries out a sequence
of operations---inserting formulas, creating pivot tables, writing and running
scripts---over a live workbook.
A task rarely has a single correct answer (``total revenue by region'' can be a
pivot table, a \Code{SUMIF} formula, or hand-entered values), and correctness
can hinge on workbook state the grid does not expose, such as formulas and
cross-sheet references; with natural-language ambiguity and model
non-determinism, ground truth is hard to
establish~\citep{aleti2023software, dobslaw2025challenges}.

Evaluation is also \emph{expensive}.
Running the agent on a \emph{task}---a workbook state paired with a
natural-language instruction---consumes tokens and time, and the cost
multiplies because the evaluation target is not the LLM alone but a
\emph{configuration}: a combination of LLM, system prompts, tool
interfaces, execution harness, and guardrails (validity or safety checks that
can block or rewrite an action), any of which can affect
outcomes, and all of which teams iterate on routinely.
As a concrete illustration of this cost, after integrating a new frontier model
into Microsoft's Excel Agent, we estimated that a single
full-benchmark run took about four hours and cost roughly \$1{,}000 in tokens at
retail prices.
Exhaustive evaluation of every candidate configuration is impractical in a
typical CI/CD setting~\citep{memon2017taming, herzig2015art}, which demands
principled test minimization~\citep{yoo2012regression}: selecting a
\emph{reduced benchmark}, a small subset of tasks that preserves the
\emph{predictive accuracy} of the full \emph{benchmark}---how well the subset's
outcomes recover the decision-relevant quantities the full benchmark would
report (e.g., pass rates, configuration rankings).

A subset chosen on purely predictive grounds, however, may concentrate in a
narrow region of the task space, dominated by a few similar task types while
whole categories of behavior go untested.
While a few ad-hoc task categorizations
exist~\citep{li2023sheetcopilot, chen2025sheetagent, li2024mimotable},
such coverage has not been defined for spreadsheet AI agents.
This gap is our starting point.

In this paper we develop a \emph{taxonomy of spreadsheet tasks} covering
properties of the spreadsheet and natural language instruction.
We build the taxonomy following the iterative method of
\citet{nickerson2013method}, grounding it in a literature review of spreadsheet
benchmarks and refining it through product team feedback at Microsoft, where
Excel Agent is being built and improved.
The taxonomy provides an externally observable structure for describing what a
benchmark covers, selecting reduced benchmarks, and interpreting failures.
We do not claim a single ``right'' taxonomy---that would overfit current
benchmarks and agents---but a reusable framework that can be adapted as the
domain evolves~\citep{beizer2003software}.

We then use the taxonomy to guide minimization.
Treating tasks as features and configurations as data points, we use greedy
forward selection with regularized regression to identify a predictive subset;
when candidates are near-tied on predictive accuracy, a taxonomy-aware
tie-breaker selects the one that maximizes diversity.
On real logs from Microsoft, when we evaluate the reduced benchmark on
configurations not used to select it (\emph{honest validation}), selecting for
coverage costs little predictive accuracy and often improves it: diversity
behaves less as a trade-off against accuracy than as a \emph{regularizer} over
greedy selection---giving up a little fit on the selection data for better
prediction on configurations not yet seen---strongest when data is noisy and
the task budget is small.
The reduced benchmarks do not, however, transfer to configurations far later in
time when trained on little recent history, which limits where taxonomy-guided minimization can be used.
Matched-budget controls---uniform random, taxonomy-stratified random,
diversity-only, and random choice within the same tolerance band---show that
the coverage gain comes from the taxonomy rather than from relaxing the
greedy objective alone, and that a predictive stage remains necessary.
Repeating the analysis on the subset of the logs with no missing outcomes,
where every configuration is scored on one common reference set, reproduces
the pattern, and taxonomy guidance again beats matched random choice within
the band under all three validation strategies, so the result is not an
artifact of imputation.

Concretely, our contributions are:
\begin{itemize}[leftmargin=*]
  \item A \emph{taxonomy of spreadsheet tasks} for evaluating AI agents,
    synthesized from the literature and industrial practice
    (Section~\ref{sec:taxonomy}).
  \item A \emph{taxonomy-guided benchmark reduction methodology} that selects a predictive \emph{reduced benchmark} and breaks near-ties by taxonomic diversity.
    On real industrial \emph{logs}---the recorded outcomes of past
    evaluations, i.e., which tasks passed or failed under each configuration---this
    tie-breaker widens taxonomy coverage with little or no loss of predictive accuracy and acts as a regularizer, as matched-budget controls and a complete-case replication confirm
    (Section~\ref{sec:evaluation} and Section~\ref{sec:test-selection}).
  \item Experience applying the taxonomy at Microsoft to audit coverage
    gaps between benchmarks and production traffic, and articulating the method's
    practical limits (Section~\ref{sec:discussion}).
\end{itemize}

\section{Related Work}
\label{sec:relwork}

We review three lines of work: methods for reducing evaluation cost, from
software-testing minimization to LLM benchmark subset selection; agent and
spreadsheet benchmarks; and how researchers define coverage for black-box systems through diversity and taxonomies.

\smallskip
\paragraph{Efficient Test and Benchmark Reduction}
\label{sec:rw:reduction}
When rerunning a full test suite on every change is too costly, software teams
cut the cost by \emph{selecting}, \emph{minimizing}, or \emph{prioritizing}
tests---running a representative subset, dropping redundant tests, or running
the most informative tests
first~\citep{yoo2012regression, elberzhager2012reducing,
hemmati2017prioritizing}---a 
pressure that long-running CI pipelines intensify~\citep{cheng2024revisiting}.
To decide which tests are worth keeping, classical methods measure
\emph{coverage}: how much of the program's internals---its statements, or
artificially injected faults---the tests exercise~\citep{hsu2009mints}.
Such code-level coverage is unavailable when a ``test'' is a natural-language
task run against a black-box system, so an externally observable
artifact---such as the requirement a test relates to---can serve as the
selection signal instead~\citep{pan2025requirements}.

The same problem arises in ML-system
testing~\citep{li2024prioritizing, dang2024test, openja2024empirical} and LLM
evaluation, where the adequacy criterion shifts from code coverage to preserving
aggregate scores or rankings.
A growing family of methods selects small static subsets (coresets) that
approximate full-suite scores or preserve model
rankings~\citep{polo2024tinybenchmarks, vivek2024anchor, zhao2024bento},
though the best subset can depend on whether it is model-agnostic or
model-aware~\citep{yuan2025tailoredbench, perlitz2024efficientbenchmarking}.

Other methods instead exploit learnable structure in model$\times$task
performance: \citet{ye2023predictable} search for a ``small-bench'' via LLM
capability predictability; \citet{zhao2025vlpredict} frame evaluation as matrix
completion; Item Response Theory estimates remaining outcomes from partial
observations by modeling item difficulty and
discrimination~\citep{cai2016irt, polo2024tinybenchmarks, li2024activeeval}; and
Active Evaluation Acquisition~\citep{li2024activeeval} and Decentralized
Arena~\citep{yin2025decentralized} choose what to evaluate next from observed
results.
The matrix-completion and active-evaluation methods spend an
\emph{observation} budget: they choose which missing configuration--task
outcomes to acquire or estimate next.
We instead fix a reusable task suite under a task budget, so the two settings
are similar in spirit but not directly comparable; the adaptive methods also
assume a missingness pattern that production logs do not have
(Section~\ref{sec:data-rep}).
We do not compare against them and claim no superiority over them.

These subsets can be fragile: prompt
sensitivity~\citep{chatterjee2024posix, polo2024prompteval} and distribution
shift~\citep{pacchiardi2024hundredinstances, zhang2025missesmark} can alter
which items are informative, motivating selection criteria that do not rely
solely on historical performance.

Our problem is closest to \emph{static subset selection under a fixed budget}:
we choose which spreadsheet tasks to run, to preserve both
decision-relevant aggregates (pass rate, configuration rankings) and structural
breadth.
Our pure predictive greedy selector ($\epsilon{=}0$, Section~\ref{sec:test-selection})
belongs to this family and represents it in our experiments; we did not run
the published instantiations because they fit on dense model$\times$task
matrices and estimate one model's score rather than rankings across
configurations.

\smallskip
\paragraph{AI Agent and Spreadsheet Benchmarks}
\label{sec:rw:agents}
Agent benchmarks span code (SWE-bench~\citep{jimenez2024swebench},
AgentBench~\citep{liu2023agentbench}), web and desktop
(WebArena~\citep{zhou2024webarena}, OSWorld~\citep{xie2024osworld}), multi-step
reasoning (GAIA~\citep{mialon2024gaia}), and API tool use
(ToolLLM~\citep{qin2024toolllm}), but none addresses \emph{selecting} a
representative subset when full evaluation is too costly, nor offers a coverage
framework for minimization.

For spreadsheet agents specifically, SheetCopilot~\citep{li2023sheetcopilot}
defines six task categories, SheetAgent~\citep{chen2025sheetagent} introduces
five broad types with 36 subtypes paired with reasoning challenges,
SpreadsheetBench~\citep{ma2024spreadsheetbench} collects 912 real forum
questions, MiMoTable~\citep{li2024mimotable} categorizes tasks by meta-operation, and InstructExcel~\citep{payan2023instructexcel} provides
crowdsourced natural-language instructions over real Excel files.
Each defines its own ad-hoc task categories; none amounts to a systematic
taxonomy designed for coverage-oriented benchmark reduction.

\smallskip
\paragraph{Black-Box Coverage via Diversity and Taxonomies}
\label{sec:rw:coverage}
When source-level coverage is unavailable---as in our black-box setting, and in
DNN testing, where neuron-coverage metrics correlate weakly with fault detection
~\citep{li2019structural, aghababaeyan2023black, gerasimou2020importance,
pei2017deepxplore, kim2019guiding}---a 
common alternative is \emph{diversity}: selecting dissimilar test cases to
broaden exploration of the fault
space~\citep{chen2010adaptive, hemmati2017prioritizing, hemmati2013achieving,
feldt2016test},
which is competitive with white-box techniques when source-level information is
inaccessible~\citep{henard2016comparing}.
Diversity often correlates positively with fault
detection~\citep{zohdinasab2021deephyperion, aghababaeyan2023black,
fahmy2021supporting},
is grounded theoretically by the observation that most faults stem from
interactions of few parameters~\citep{kuhn2004software}, is operationalized
by multi-objective and distance-aware methods in black-box
DNN testing~\citep{aghababaeyan2024deepgd, li2024distance}, and has been
combined with fault-proneness estimates to prioritize
tests~\citep{mahdieh2022test}; \citet{elgendy2025systematic} map
diversity-based testing.

Diversity, however, raises the question: \emph{with respect to what?}
A common answer is a \emph{taxonomy}---a set of dimensions serving as
an operational coverage scheme.
Autonomous-vehicle testing derives scenario taxonomies via clustering or
driving-dimension
characterization (e.g., \citep{neelofar2024towards}); 
LLM testing explores model knowledge boundaries through knowledge
graphs~\citep{rajan2024knowledge}; and web testing uses defect taxonomies to
build smaller, more effective
suites~\citep{marchetto2009empirical, rajaram2019taxonomy}. 

\section{Knowledge Gap and Problem Framing}
\label{sec:framing}

Classical minimization defines test adequacy through structural
coverage---statement or mutation coverage---which cannot be measured for an LLM-based agent.
Coreset methods for LLM benchmarks select on historical performance
correlations, with no notion of which behaviors a subset exercises, and turn
fragile once those correlations break down under prompt sensitivity or drift
over time.
Diversity is the standard black-box substitute for coverage, but it presupposes
dimensions along which to be diverse---and existing spreadsheet and agent
benchmarks supply only ad-hoc task categories, not a systematic taxonomy built
for coverage-oriented selection.
Pairing a domain-grounded taxonomy with a predictive subset-selection procedure
is our response---but it requires a specific framing of the selection
problem.

\smallskip
\paragraph{Why a Reduced Benchmark}
Production agent teams continuously compare candidate configurations across a
large design space: cheaper or faster LLMs that may underperform
slightly but win on cost or latency; fine-tuned variants whose quality is
uncertain until measured; and harness-only changes that keep the model
fixed---alternative spreadsheet representations, tool schemas, retrieval
settings, safety policies, or new tool integrations.
Each needs a full-benchmark score before it can be compared with the current default, yet a typical CI/CD cycle exercises thousands of tasks under rate
limits.
A \emph{reduced benchmark}---a small, fixed subset whose outcomes approximate
the full-benchmark score for any new configuration---trades a one-time curation
cost for a recurring saving.

\smallskip
\paragraph{Why this works}
The framing is well-posed when configurations vary along a few latent
capabilities and tasks are redundant within them: a small set of informative
tasks with learned weights can then recover the full-benchmark score
(App.~\ref{app:toy-example} gives a stylized
example).

\smallskip
\paragraph{Tasks as Features, Configurations as Datapoints}
This structure---a few latent capabilities with many redundant tasks---suggests
the general framing.
A uniform random sample of tasks would suffice only if every task correlated
equally well with the full-benchmark score, but tasks are heterogeneous---they
probe different operations and difficulty levels---so a small random sample can
over-represent one region and predict poorly.
We therefore choose tasks \emph{for how well they predict that score}, casting
the choice as supervised learning: \emph{tasks are features} and
\emph{configurations are datapoints}.
Each past configuration has been run against many tasks, giving a vector of
pass/fail outcomes whose average---its full-benchmark score---is the label, and
selecting a reduced benchmark is feature selection.

This requires a pool of past configurations with known full-benchmark scores,
which may seem self-defeating if running the full benchmark is too costly.
The resolution is temporal: teams do run the full benchmark periodically---to
certify releases, set baselines, or vet major model upgrades---and these runs
accumulate over a project's lifetime.
Whether the backlog is large and varied enough to support selection is an
empirical question we examine using logs from Microsoft
(Section~\ref{sec:test-selection}).

These logs are also the right \emph{kind} of training data: they do not require new data collection and arise from realistic axes of variation---model swaps,
prompt edits, retrieval changes.
What we extract from them is a property of the \emph{tasks}, not of any single
configuration: which tasks are informative about aggregate behavior and which
are redundant.
A configuration that will never be deployed is therefore still useful
training data, and the signal depends only on relative \emph{differences}
across configurations, easier to estimate than any absolute score.
When extrapolation across model generations is a concern, the training pool
can be restricted to a recent slice.

In the dominant coreset framing, one picks a subset to estimate \emph{one fixed model's} score; we instead pick a subset to compare and rank \emph{many candidate
configurations}.

\smallskip
\paragraph{From framing to method}
This framing imposes two requirements.
First, predictive accuracy alone is not enough: a reduced benchmark chosen
purely to maximize $R^2$---the fraction of variance in the full-benchmark score
that a reduced benchmark's predictions explain---can cluster in narrow
regions of the task space, so we
need a \emph{coverage criterion} meaningful without source-level signals---a
domain-grounded taxonomy answering ``with respect to what''
(Section~\ref{sec:taxonomy}).
Second, we need a \emph{predictive selection algorithm} that picks a small,
weighted benchmark under that taxonomy, whose design decisions we test empirically on real logs (Section~\ref{sec:test-selection}).

\section{The Taxonomy of Spreadsheet Tasks}
\label{sec:taxonomy}

This section presents a taxonomy of externally observable properties of the
workbook and the task instruction---the dimensions over which we operationalize
coverage.
We describe first how we identified its dimensions (Section~\ref{ssec:methodology})
and then the result (Section~\ref{ssec:taxonomy-table}).

A \textit{workbook} contains one or more \textit{spreadsheets} (sheets, or
``tabs''), each a two-dimensional grid of cells, and a spreadsheet contains one
or more \textit{tables}---contiguous regions holding a coherent unit of data
(e.g., a header row with its records).
We present the dimensions primarily at the \textit{table} level, the finest
granularity at which most are expressed and the natural scope of many tasks
(lookup, aggregation, cleaning).\footnote{The spreadsheet- and workbook-level
  rows are mostly \emph{aggregations} of table-level dimensions: a dimension
  defined at the table level (e.g., \emph{Table size}) induces a corresponding
  workbook-level characterization by composition (e.g., \emph{Size}).
  Table~\ref{tab:taxonomy} lists only a few such workbook-level dimensions
  (\emph{Size}, \emph{Structural complexity}, \emph{Domain/Topic}) to avoid
clutter.}

\begin{table}[t]
  \centering
  \scriptsize
  \caption{Our taxonomy dimensions and the literature informing them. Each
    \textbf{dimension of variation} of the workbook (or its components) or of
  the task instructions has one or more \textbf{concrete measures}.}
  \label{tab:taxonomy}
  \setlength\extrarowheight{0.5mm}
  \setlength{\tabcolsep}{2pt}
  \begin{tabular}{L{1.5cm}L{2.5cm}L{4.5cm}}
    \toprule
    Dimension & Description & Example Measures [Sources] \\
    \midrule
    \multicolumn{3}{l}{\textit{\textbf{Properties of a Table}}} \\
    Table size & Table grid dimensions & Number of rows \& columns
    \citep{ma2024spreadsheetbench, cui2025nlctables, wu2025tablebench,
    chen2024sheetagent, chen2025sheetagent} \\
    Vocabulary size & Size of vocabulary & Number of unique values
    \citep{eggert2023tablib} \\
    Header type & The structure of the header row/column & A categorical
    label, e.g., \textit{simple} (single-row or single-column header) vs
    \textit{complex} (e.g., hierarchical) \citep{li2024mimotable,
    ma2024spreadsheetbench, wu2025realhitbench} \\
    Structure & The structure of the table & A categorical label, e.g.,
    \textit{flat} vs \textit{hierarchical} \citep{sui2024table,
    wu2025realhitbench, lu2025large, ma2024spreadsheetbench} \\
    Data types & The types of data contained in the table & A list, e.g.,
    \textit{string}, \textit{numeric}, \textit{date} \citep{eggert2023tablib,
    cui2025nlctables, lu2025large} \\
    Data cleanliness & The cleanliness of the data in the table (e.g.,
    missing entries, inconsistently formatted values) & A categorical label,
    e.g., \textit{clean} vs \textit{noisy} \citep{singha2023tabular, li2024table} \\
    Language & The primary language used in the table & A categorical label,
    e.g., \textit{English} \citep{eggert2023tablib, li2024mimotable} \\
    \addlinespace[\TableLineExtraSpace]
    \multicolumn{3}{l}{\textit{\textbf{Properties of a Spreadsheet or the
    Workbook}}}                \\
    Size & The size of the parent (spreadsheet/workbook) as a set of
    components (tables/sheets) or the file as a whole. & Number of sheets
    \citep{chen2024sheetagent, chen2025sheetagent, ma2024spreadsheetbench,
    li2024mimotable, wu2025mmqa} \newline \newline Number of tables
    \citep{ma2024spreadsheetbench, li2024mimotable, wu2025mmqa} \newline
    \newline File size on disk \citep{payan2023instructexcel, wu2025tablebench} \\
    Structural complexity & The overall structural complexity of the parent.
    & A derived rating that \emph{aggregates} table-level structure and size
    signals, e.g., summarized by their principal components \citep{li2024mimotable} \\
    Domain / Topic & The domain or topic of the workbook & A categorical
    label, e.g., \textit{architecture}, \textit{finance}, \textit{education},
    \textit{accounting} \citep{he2022anameta, hulsebos2023gittables,
      eggert2023tablib, li2024mimotable, wu2025realhitbench, cui2025nlctables,
      ashury2025mighty, wu2025tablebench, chen2024sheetagent,
    chen2025sheetagent} \newline \newline NAICS industry sector [this work]
    \newline \newline Workbook primary intent [this work] \\
    \addlinespace[\TableLineExtraSpace]
    \multicolumn{3}{l}{\textit{\textbf{Properties of the Task Instructions
    (and Task As a Whole)}}}                \\
    Clarity & The clarity of the natural language instructions & A
    categorical label, e.g., \textit{clear} vs \textit{ambiguous}
    \citep{chen2024sheetagent, chen2025sheetagent, lu2025large} \newline
    \newline Instruction specificity rated on a 1--5 rubric [this work] \\
    Operations & Specific operations that are part of the instructions, with
    or without their objects & List of query verb-noun phrases (actions +
    objects) \citep{chen2024sheetagent, chen2025sheetagent,
    ma2024spreadsheetbench} \newline \newline List of meta operations (e.g.,
      \textit{lookup}, \textit{edit}, \textit{calculate}, \textit{compare},
    \textit{visualize}, \textit{reason}) \citep{payan2023instructexcel,
    li2024mimotable, chen2024sheetagent, chen2025sheetagent} \\
    Task type & The type of the task to be performed & A categorical label,
    e.g., \textit{table Q\&A}, \textit{table manipulation}, \textit{data
    analysis} \citep{li2023sheetcopilot, zhang2023tablellama,
      li2024mimotable, li2024table, zhang2024tablellm, wu2025realhitbench,
      xing2025mmtu, ashury2025mighty, yi2025tablepilot, wu2025tablebench,
    lu2025large, chen2024sheetagent, chen2025sheetagent} \newline \newline
    Action type: \textit{ask} vs.\ \textit{act} vs.\ \textit{unknown}~[this
    work] \newline \newline Multiple operationalizations (thematic analysis,
    capability-, and intent-based) [this work] \\
    Task difficulty & Task horizon or complexity & Number of
    operations/reasoning steps required \citep{chen2024sheetagent,
    chen2025sheetagent} \newline \newline
    A weighted average of the difficulties of the task's meta operations
    (e.g., lookup and edit can be considered the easiest, while reasoning can
    be considered the hardest) \citep{li2024mimotable, chen2024sheetagent,
    chen2025sheetagent} \\
    LLM capabilities & The capabilities expected of the language model in
    handling the task & E.g., \textit{error detection}, \textit{symbolic
    reasoning}, \textit{domain understanding}, \textit{numerical
    abstraction}, \textit{explanation} \citep{thorne2025large,
      ashury2025mighty, li2025treb, lu2025large, chen2024sheetagent,
    chen2025sheetagent} \\
    \bottomrule
  \end{tabular}
\end{table}

\subsection{Methodology for Building our Taxonomy}
\label{ssec:methodology}

Building a taxonomy requires choosing both \emph{which dimensions of variation}
to include and, for each, \emph{how to measure it}---the concrete measure or
measures that assign a task a value along that dimension.
These decisions are interdependent: the right dimensions emerge only from
examining real tasks, yet examining tasks presupposes at least a tentative
scheme.
We resolve this circularity by following the iterative taxonomy-development
method of \citet{nickerson2013method}, which alternates \emph{deductive}
(top-down) cycles---deriving candidate dimensions from theory and testing them
against data---with \emph{inductive} (bottom-up) cycles---examining data to
find dimensions not yet in the scheme---until a set of ending conditions is
met.
\citet{nickerson2013method} give two kinds of ending condition.
The \emph{objective} ones hold when every object examined has been classified
and a full iteration does not add, merge, or split any dimension or value; the
\emph{subjective} ones hold when the taxonomy is concise, robust,
comprehensive, extendible, and explanatory.
A single guiding criterion governs what the taxonomy includes, and every
dimension must serve it.

\smallskip
\paragraph{Instantiation}
We apply this method to AI spreadsheet agents.
Our objects of interest are \emph{spreadsheet tasks}, each defined by a workbook
state and a natural-language instruction.
Our guiding criterion is that every dimension capture an
\emph{externally observable} property of the workbook or instruction---one
relevant for describing, selecting, and interpreting agent evaluations---since
agent internals are inaccessible in our black-box setting.
Both cycle types require concrete tasks, so we need a task corpus to iterate
over.

This corpus is a convenience sample drawn from one popular public benchmark,
SpreadsheetBench~\citep{ma2024spreadsheetbench}, and an internal benchmark under
development at Microsoft.
Inspecting every public benchmark (Section~\ref{sec:rw:agents}) was impractical;
combining a public and an internal source lets the taxonomy reflect both the
academic view of spreadsheet tasks and the workloads of a production agent.

\smallskip
\paragraph{Identifying dimensions}
In deductive (conceptual-to-empirical) cycles we started from theory: drawing
on the benchmarks surveyed in Section~\ref{sec:rw:agents}, we proposed candidate
dimensions, applied them to corpus tasks, and revised them when a dimension did
not vary (an early \emph{file format} candidate: nearly every task is an
\texttt{.xlsx} workbook), was redundant (a per-table ``structural complexity''
  rating that restated \emph{Header type}, \emph{Structure}, and
\emph{Table size}, kept only as an aggregate workbook property), or conflated
separable axes (a single ``type'' dimension, split into \emph{Task type} and
the \emph{Operations} it comprises)~\citep{marchetto2009empirical}.

In inductive (empirical-to-conceptual) cycles we started from data,
open-coding corpus tasks to find recurring properties not yet captured and
grouping them into new or revised dimensions; \emph{data cleanliness} emerged
this way, after we repeatedly found tables with missing values that our theory-derived dimensions did not capture.
Open coding also yielded candidate measures and value sets for each
dimension.
We iterated through both cycle types, with rounds of discussion between two
researchers~\citep{marchetto2009empirical, lupetti2024making}, until the
taxonomy stabilized and met our ending conditions.

\smallskip
\paragraph{Choosing concrete measures and stopping}
For each dimension we then chose one or more \emph{concrete measures}, using
programmatic extraction for objective workbook structure and LLM-assisted coding
for interpretive task or instruction properties.
We allowed multiple measures when a dimension had several useful measures: for
example, \emph{Task type} can be grounded in instruction themes, required agent
capabilities, or workbook intent, each with different tradeoffs.
We stopped when every corpus task could be classified by every dimension and the
taxonomy satisfied the subjective ending conditions above (concise, robust,
comprehensive for our scope, extendible, and explanatory).
App.~\ref{app:taxonomy-method-details} gives the full coding and termination
details.

\subsection{The Resulting Taxonomy}
\label{ssec:taxonomy-table}

Table~\ref{tab:taxonomy} summarizes the taxonomy: at the table level it
captures size, vocabulary, headers, structure, data types, cleanliness, and
language; at the spreadsheet/workbook level scale, structural complexity, and
domain; and at the task level clarity, operations, task type, difficulty, and
required LLM capabilities.
Because the table lists only \emph{representative} measures and values,
App.~\ref{app:taxonomy-values} enumerates the complete value sets and
App.~\ref{app:taxonomy-examples} gives SpreadsheetBench examples for each
tier.

\begin{figure*}[t]
  \centering
  \begin{minipage}[c]{0.46\linewidth}
    \centering
    \subfigure[Task
    type.]{\includegraphics[width=0.74\linewidth]{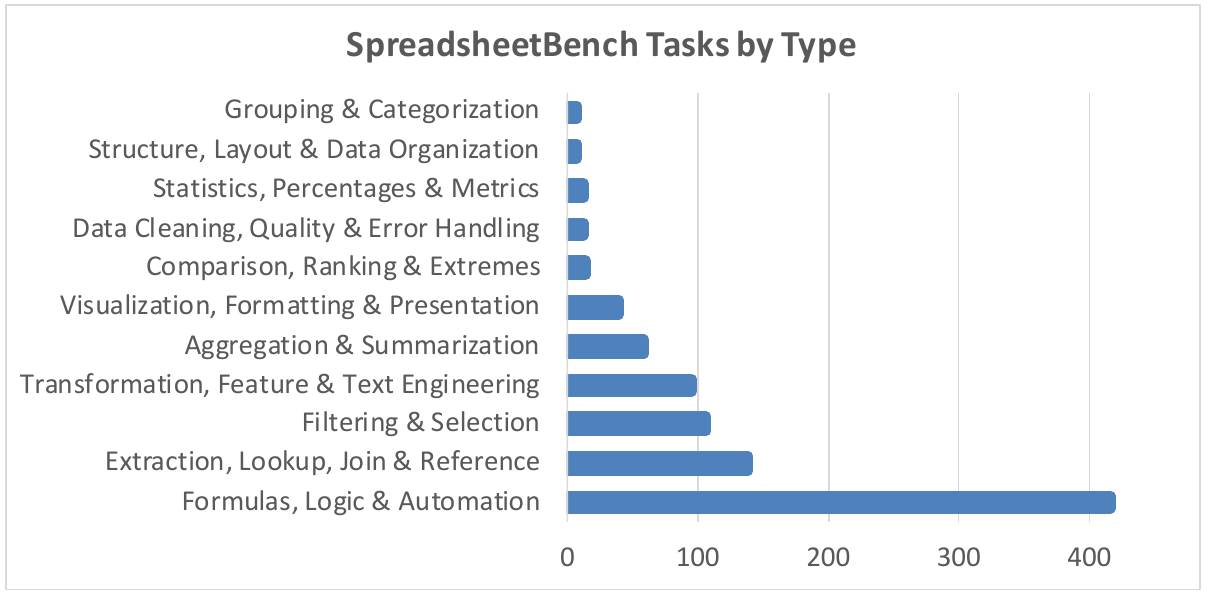}}\\[4pt]
    \subfigure[Steps implied (task
    horizon).]{\includegraphics[width=0.485\linewidth]{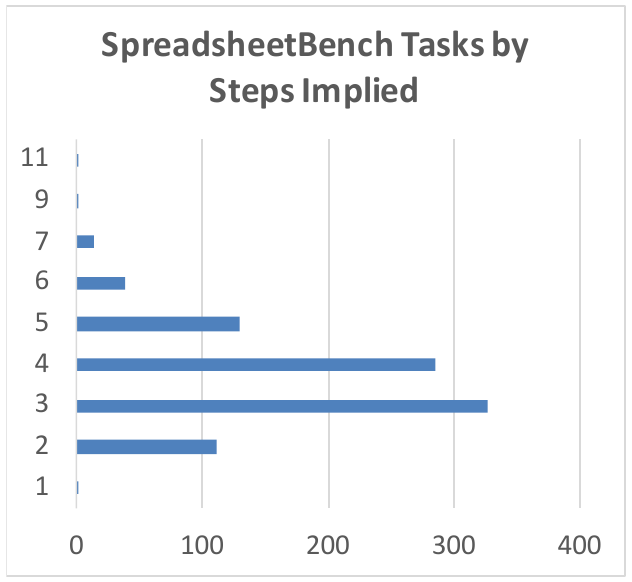}}\hfill
    \subfigure[Instruction
    specificity.]{\includegraphics[width=0.485\linewidth]{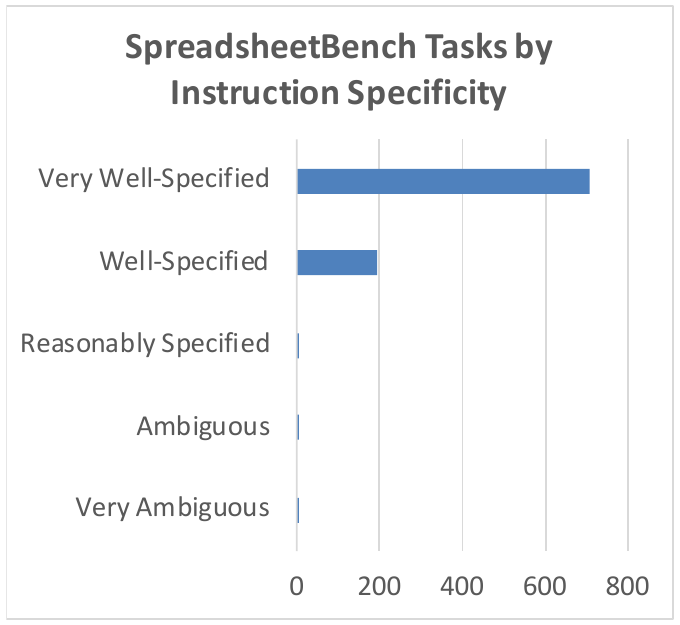}}
  \end{minipage}\hfill
  \begin{minipage}[c]{0.52\linewidth}
    \centering
    \subfigure[Workbook intent
    (domain).]{\includegraphics[width=0.74\linewidth]{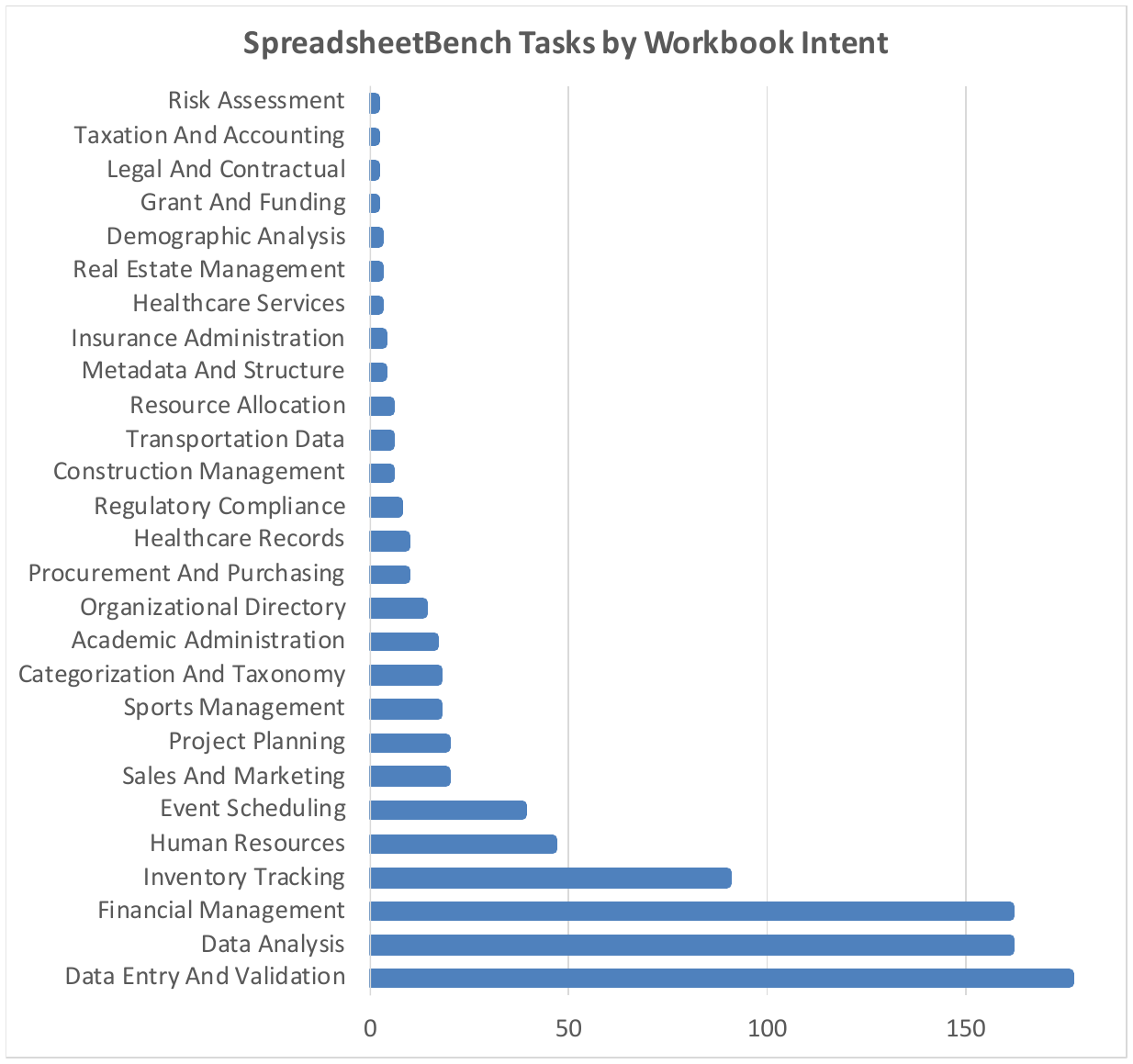}}
  \end{minipage}
  \caption{SpreadsheetBench tasks across four taxonomy dimensions: a few
    categories dominate while many are sparsely populated, so the dimensions
    discriminate among tasks (descriptive validity) and benchmark coverage of
  the taxonomy is highly uneven.}
  \label{fig:sb-distributions}
\end{figure*}

\subsection{Validity of the Taxonomy}
\label{sec:evaluation}

Our main evaluation---whether the taxonomy enables effective
\emph{benchmark reduction}---is the subject of Section~\ref{sec:test-selection}.
First we establish a precondition, \emph{descriptive validity}: that the
dimensions discriminate among tasks, partitioning a real benchmark into
substantively different groups.
We illustrate this on SpreadsheetBench~\citep{ma2024spreadsheetbench}, which we
annotated end-to-end following the labeling procedure of
Section~\ref{ssec:methodology}.
Concretely, we compute objective measures such as workbook size
from the workbook metadata, and use an LLM (GPT-4.1)
for subjective measures such as task type and instruction specificity,
prompted with the measure's definition and value set as a rubric.

One author spot-checked the LLM labels by hand-coding a sample, concentrating
on the most subjective dimensions, task type and workbook intent, and a second
author reviewed the disagreements.
The checks did not reveal any major concerns beyond the subjectivity one expects from these dimensions.
Disagreements arose with sparse workbooks, whose few filled cells do not reveal the domain or purpose, and with instructions that combine operations, where
more than one task type fits: for example, highlighting each row's best value is both
\emph{Comparison, Ranking \& Extremes} and
\emph{Visualization, Formatting \& Presentation}, and the LLM tends to label
anything that fits nowhere else as \emph{Formulas, Logic \& Automation}, which
may inflate that category's count in Fig.~\ref{fig:sb-distributions}(a).
In these cases neither the human nor the LLM was clearly right, and a more
detailed coding guide would be needed to resolve them.
The distributions below show that the dimensions separate tasks into different
groups (descriptive validity); they do not show that every label is right, and labels affect
selection only through the tie-break (Section~\ref{sec:discussion}).

Figure~\ref{fig:sb-distributions} shows four of the most interpretable
dimensions.
\emph{Task type} (a) is heavy-tailed: \emph{Formulas, Logic \& Automation}
alone accounts for over 400 tasks (a count that the catch-all labeling noted
above may inflate); \emph{Extraction, Lookup, Join \& Reference} and
\emph{Filtering \& Selection} follow in the low hundreds; and
\emph{Grouping \& Categorization} and
\emph{Structure, Layout \& Data Organization} appear only a handful of times.
Figure~\ref{fig:sb-distributions}(b) shows \emph{steps implied}, our proxy for
task horizon (\emph{task difficulty} in Section~\ref{ssec:taxonomy-table}): unimodal
but right-skewed, peaking at three to four steps, with a thin tail of
long-horizon tasks demanding seven or more.
\emph{Instruction specificity} (c; the \emph{clarity} dimension) is
concentrated at the high end: nearly all instructions are \emph{Well-} or
\emph{Very Well-Specified}, so SpreadsheetBench barely probes the underspecified
regime.
\emph{Workbook intent} (d; domain) spans more than two dozen domains but
clusters in \emph{Data Entry \& Validation}, \emph{Financial Management}, and
\emph{Data Analysis}, with a long tail (e.g., \emph{Risk Assessment},
\emph{Legal \& Contractual}) represented by a few tasks each.

Each dimension, then, is populated across multiple, clearly distinguishable
levels, so the dimensions discriminate among tasks; catch-all labeling
inflates one task-type category, but the remaining task types and the other
three dimensions show the same spread.
The skew also affects evaluation, even if the largest category is
overstated: because a few task types and domains dominate and many are
represented by a handful of tasks, uniform random sampling over-represents the
head and under-samples the tail, so benchmark size does not guarantee taxonomy
coverage---which motivates the taxonomy-aware
reduction we study next.

\section{Taxonomy-Driven Benchmark Reduction}
\label{sec:test-selection}

We need two things (Section~\ref{sec:framing}): a coverage criterion grounded
in the spreadsheet taxonomy (Section~\ref{sec:taxonomy}) and a predictive selector
that uses it.
This section makes both concrete and evaluates them on real industrial
\emph{logs} (the recorded pass/fail outcomes of past evaluations).
Our method is to represent the logs as a configuration-by-task
matrix, run a greedy predictive selector over it, and break ties in predictive
accuracy by taxonomic diversity.
Exact ties are rare, so relaxing ``tie'' to within $\epsilon$ of the best
predictor widens the band of contenders and increases diversity, but might cost
predictive accuracy if the looser contenders are weaker predictors.
We answer four research questions:
\begin{itemize}[leftmargin=*,itemsep=1pt,topsep=2pt]
  \item[\textbf{RQ1}] How much predictive accuracy does selecting for
    taxonomic coverage cost? (Section~\ref{sec:results-reg})
  \item[\textbf{RQ2}] When does coverage help most? (Section~\ref{sec:results-regime})
  \item[\textbf{RQ3}] Does the reduced benchmark survive \emph{temporal
    shift}---does it stay predictive as configurations evolve
    over time? (Section~\ref{sec:results-hierarchy})
  \item[\textbf{RQ4}] Is the benefit attributable to the taxonomy itself, and
    does it survive when every configuration is scored on one common, fully
    observed reference task set? (Section~\ref{sec:results-controls})
\end{itemize}
We study one design choice experimentally---how hard to push for diversity
($\epsilon$)---along with two axes that determine when it helps (the task budget
$k$ and a data-quality filter); how to impute missing entries and weight tasks
we decide by reasoned argument below.

\subsection{Data Representation}
\label{sec:data-rep}
We arrange the logs into a configuration-by-task matrix
$X\in\mathbb{R}^{n\times p}$ ($n=|\mathcal{C}|$ configurations,
$p=|\mathcal{T}|$ tasks), with $X_{c,t}=r_{c,t}$ where observed and missing
otherwise.
Missingness is common---partial runs across CI cycles,
skipped or rerun tasks, inconsistent storage as infrastructure evolves---so $X$
is sparse and non-uniform.\footnote{We mean-impute each missing entry by its
  per-task mean
  and drop all-missing and zero-variance columns, giving the dense feature
  space $\tilde{X}$ on which greedy selection and Ridge operate
(App.~\ref{app:imputation}).}
Each row of $X$ is one recorded benchmark execution; several executions can
share a treatment definition (e.g., reruns of one prompt variant), so rows are
not fully independent, a point we return to in Section~\ref{sec:sweep-grid} and
Section~\ref{sec:discussion}.

We treat one filter as an experimental lever: each configuration's
\emph{observed-outcome rate}---the fraction of its tasks actually
observed---must meet a threshold \texttt{min\_completion\_rate}; lowering it allows more but noisier configurations, and sweeping it probes sensitivity to
data noise (Section~\ref{sec:results-regime}).

\subsection{Predictive Minimization via Greedy Forward Selection}
Our prediction target is each configuration's full-benchmark score $y_c$---its
mean outcome over the tasks \emph{observed for that configuration},
$\mathcal{T}_c$ (App.~\ref{app:formalization})---collected into
$y\in\mathbb{R}^n$, which we predict from a small subset of tasks $S$ ($|S|=k$)
together with a weighting over them.
Because $\mathcal{T}_c$ differs across configurations, this target is
configuration-specific: a difference in $y_c$ can reflect which tasks were run,
and how hard they were, as well as how well the configuration did.
We keep it for the principal analysis because it is the target the logs support,
and we test in Section~\ref{sec:results-controls} whether the conclusions hold when
all configurations share one fully observed task set.
We fix the weighting first, since it also scores candidates during selection.

\smallskip
\paragraph{Weighting via Ridge regression}
Each selected task has a learned weight, and we predict a configuration's
full-benchmark score as the weighted sum of its outcomes on those tasks.
With many candidate tasks, few configurations, and many near-identical tasks,
least squares is unstable, so we use \emph{Ridge regression}, which keeps the
weights small and stable through an $\ell_2$ penalty ($\alpha=0.01$; objective
and closed form in App.~\ref{app:ridge}).
We interpret the weights $\hat{\beta}(S)$ as a stable recipe for combining
task outcomes; they do not explain why individual tasks pass or fail.

\smallskip
\paragraph{Selection via greedy forward selection}
Exact subset selection is combinatorial, so we build $S$ greedily: from
$S_0=\emptyset$, each step adds the task that most improves the Ridge fit,
scored by inner cross-validated $R^2$ (5-fold, shuffled, fixed seed), until
$|S|=k$ or the inner $R^2$ reaches a convergence threshold; this yields a
ranked, inspectable list (App.~\ref{app:greedy}).

\subsection{Taxonomy-Guided Diversity}
Since greedy selection only optimizes predictive accuracy,
we use the taxonomy to inject a coverage objective into selection, breaking
near-ties by \emph{taxonomic diversity}.

\smallskip
\paragraph{Taxonomy diversity score}
We score how well a reduced benchmark $S$ spans the taxonomy of
Section~\ref{ssec:taxonomy-table} by the unweighted mean of per-dimension coverage
scores,
\begin{equation}
  \mathrm{Div}(S) = \frac{1}{|\mathcal{D}|} \sum_{j\in\mathcal{D}} \mathrm{Div}_j(S),
\end{equation}
where $\mathcal{D}$ is every taxonomy dimension of Table~\ref{tab:taxonomy}
except \emph{data cleanliness} (not operationalized), and each
$\mathrm{Div}_j(S)\in[0,1]$ measures coverage of dimension $j$ by a form suited
to its measure type---range coverage, normalized entropy, or label coverage
(full per-dimension definitions in App.~\ref{app:diversity}).

\smallskip
\paragraph{Diversity tie-breaker via an $\epsilon$-tolerance band}
Exact ties in inner cross-validation (CV) $R^2$ are rare, but at each greedy
step many candidates
are \emph{nearly} tied.
We use these near-ties: among all candidates within $\epsilon$
of the best $R^2$,
we add the one that most increases the diversity score $\mathrm{Div}(\cdot)$:
\begin{equation}
  B_\epsilon(S_{t-1}) = \{q\notin S_{t-1} :
  R^2_{\mathrm{innerCV}}(S_{t-1}\cup\{q\}) \ge R^2_{\mathrm{best}} - \epsilon\},
\end{equation}
\begin{equation}
  q^* = \arg\max_{q\in B_\epsilon(S_{t-1})} \; \mathrm{Div}(S_{t-1}\cup\{q\}).
\end{equation}
Predictive accuracy stays primary; raising $\epsilon$ widens the band,
trading up to $\epsilon$ of inner-CV $R^2$ for coverage
(Section~\ref{sec:sweep-grid}, Section~\ref{sec:results-reg}).

\subsection{Measuring Generalization Without Selection Bias}
\label{sec:eval-setup}
Because greedy selection optimizes an inner cross-validation objective,
evaluating the reduced benchmark on the same configurations would
be biased by selection.
We therefore wrap the selection-and-fit pipeline in an outer validation loop:
each outer split holds out configurations for judging, and the rest are
used for greedy selection, inner 5-fold CV, and Ridge fitting.
This yields an \emph{honest} $R^2$ free of selection bias in the sense of
\citet{cawley2010overfitting}; for comparison, we also report an
\emph{optimistic} $R^2$ that selects tasks once on all data and therefore
leaks selection into
the held-out folds.
We use three outer strategies---shuffled Nested $K$-Fold, Walk-Forward
Expanding, and Walk-Forward Sliding---defined in App.~\ref{app:splits}.
Their hierarchy separates contemporaneous generalization from temporal
generalization: the K-Fold/Expanding gap isolates temporal leakage, while the
Expanding/Sliding gap isolates the cost of discarding older training
data.

\begin{figure*}[t]
  \centering
  \subfigure[Honest weighted-Ridge
  $R^2$]{\includegraphics[width=0.32\linewidth]{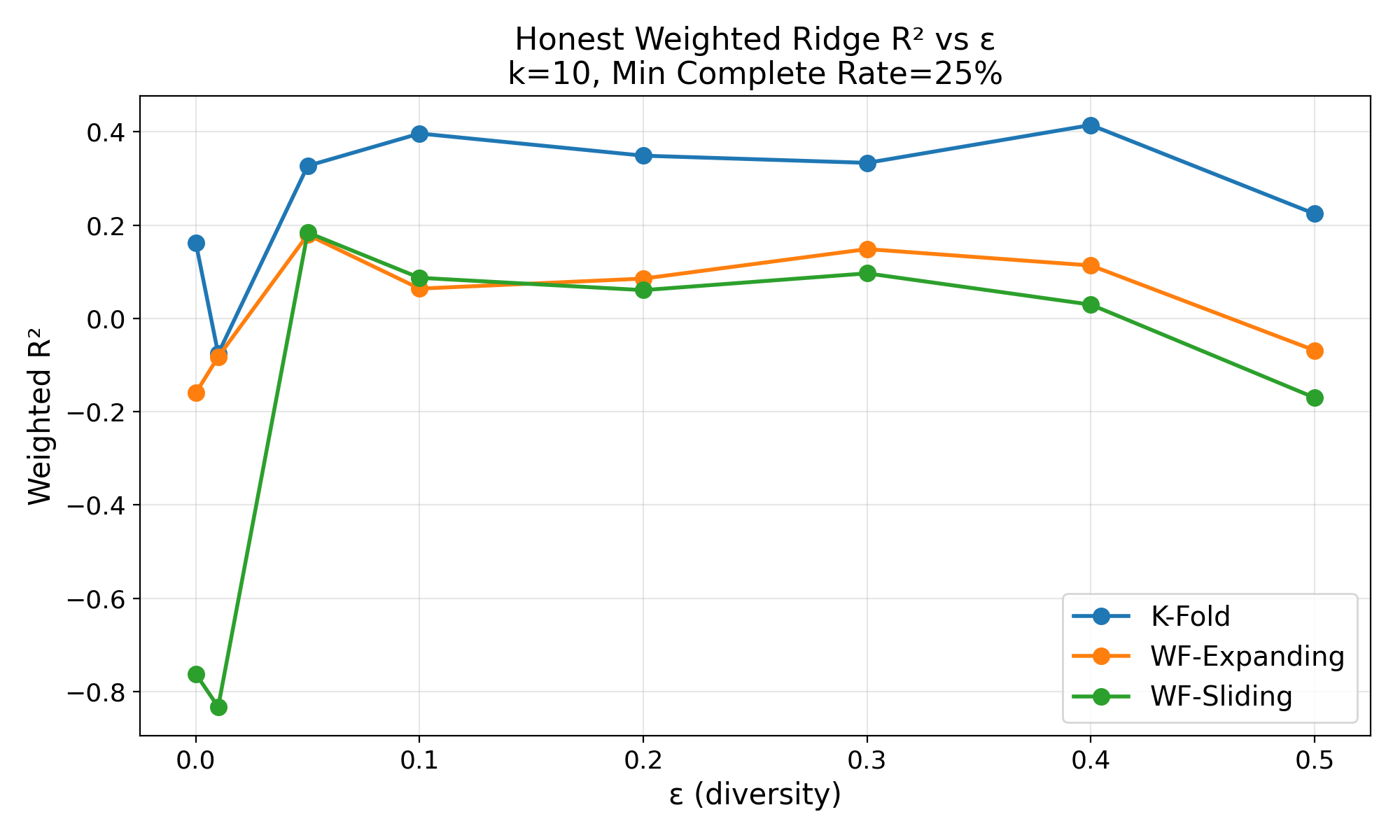}}\hfill
  \subfigure[Optimistic
  $R^2$]{\includegraphics[width=0.32\linewidth]{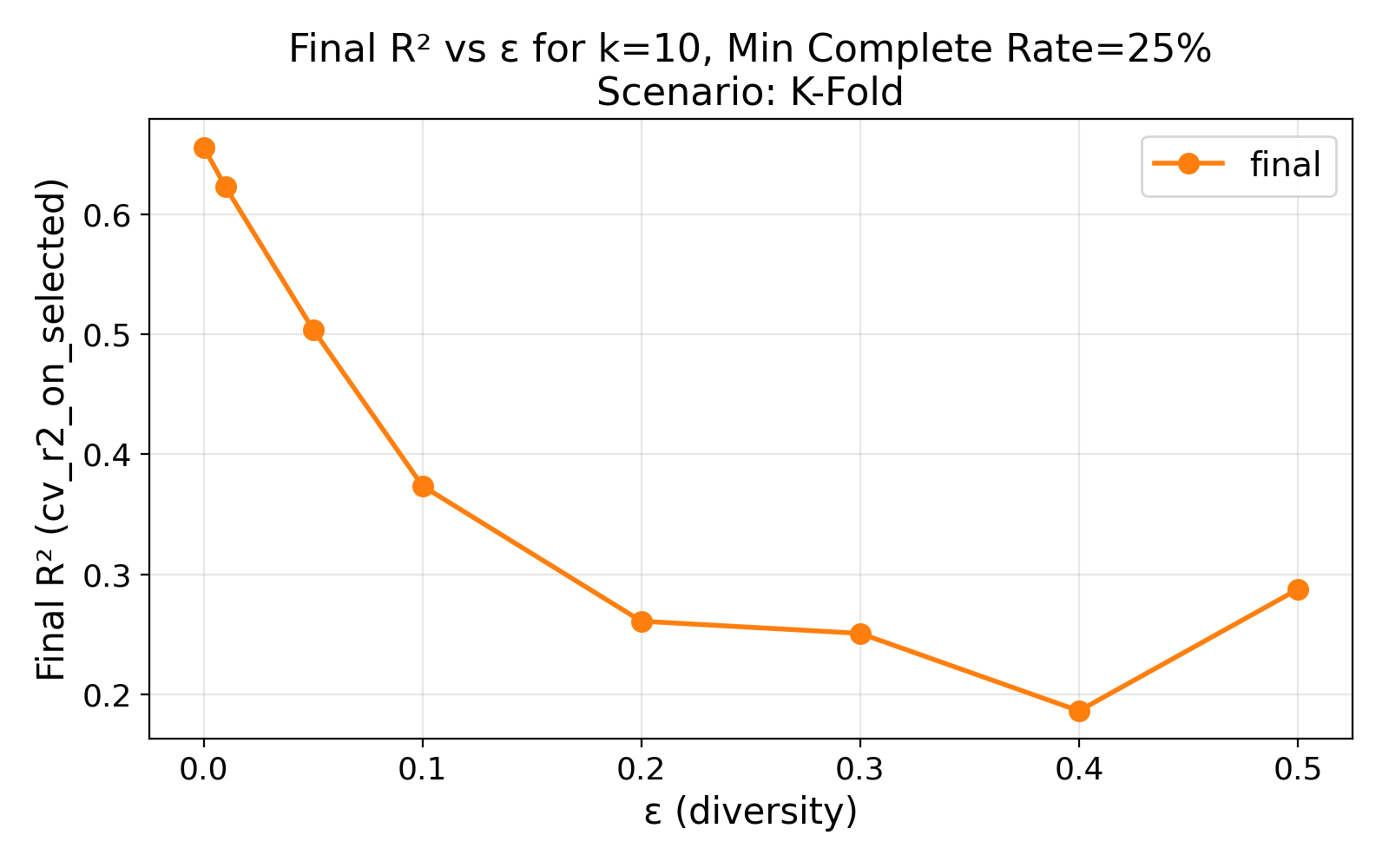}\label{fig:final-r2-by-method}}\hfill
  \subfigure[Taxonomy diversity
  $\mathrm{Div}(S)$]{\includegraphics[width=0.32\linewidth]{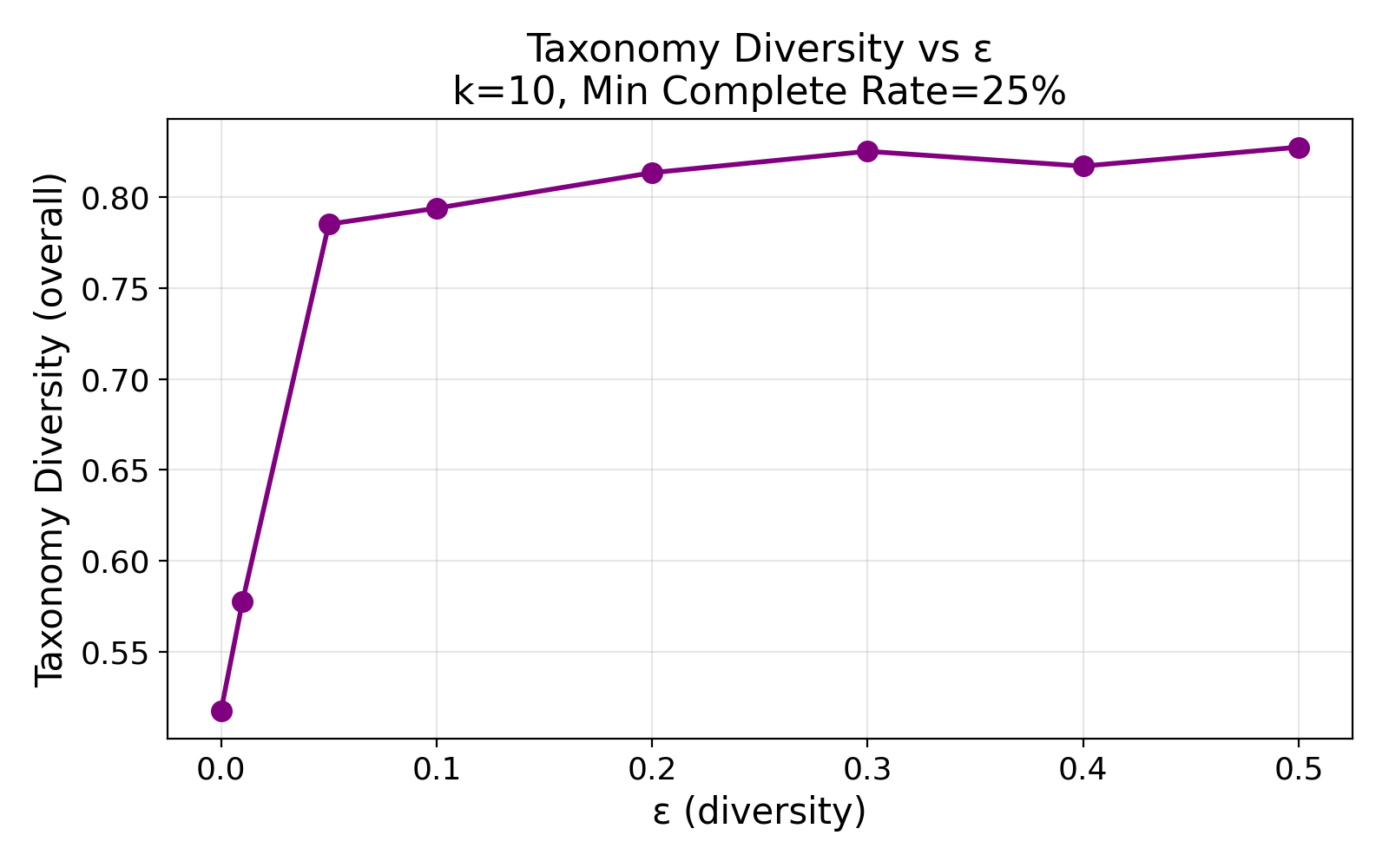}\label{fig:div-by-method}}
  \caption{All at $k{=}10$, \texttt{min\_completion\_rate}$=25\%$,
    vs.\ $\epsilon$. (a) Honest weighted-Ridge $R^2$ for the three outer-loop
    strategies: K-Fold $>$ Expanding $>$ Sliding throughout, each curve
    non-monotone, rising at small-to-moderate $\epsilon$ before falling. (b)
    Optimistic $R^2$ (independent of the outer strategy), nearly
    monotone-decreasing. (c) Taxonomy diversity $\mathrm{Div}(S)$, rising from
  pure greedy selection ($\epsilon{=}0$) and saturating by $\epsilon\approx0.1$--$0.2$.}
  \label{fig:r2-weighted-by-method}
\end{figure*}

\subsection{Data and Sweep Grid}
\label{sec:sweep-grid}
We evaluate on benchmark logs
with configurations filtered to a single production-candidate configuration
family, yielding $n=67$ configurations (126 execution IDs, since some
configurations were executed more than once) and $p=869$ candidate tasks
after dropping all-missing and zero-variance
columns.
We sweep the full grid
$\epsilon \in \{0, 0.01, 0.05, 0.1, 0.2, 0.3, 0.4, 0.5\}$,
$k \in \{7, 10, 15, 20, 25, 30\}$, and \texttt{min\_completion\_rate}
$\in \{1\%, 5\%, 10\%, 15\%, 20\%, 25\%, 30\%\}$ (the completion-rate filter
defined in Section~\ref{sec:data-rep}) across the three outer-loop strategies above,
with Ridge ($\alpha=0.01$) and 5-fold inner CV as defined above throughout.
RQ1--RQ3 each take up one design question, drawing on the representative grid
cell $k=10$, \texttt{min\_completion\_rate}$=25\%$
together with the full $k \times$ \texttt{min\_completion\_rate} grids;
throughout, \emph{honest} $R^2$ is the out-of-fold measure defined above and
\emph{optimistic} $R^2$ its selection-biased foil.

\smallskip
\paragraph{Matched-budget controls (RQ4)}
To isolate what taxonomy guidance contributes, we compare it at the same budget
$k{=}10$ against five controls:
(i)~pure predictive greedy selection ($\epsilon{=}0$);
(ii)~\emph{$\epsilon$-band random}, which forms the same near-tie band
$B_\epsilon$ but picks a member uniformly at random instead of by
diversity---the control that separates the taxonomy's contribution from
relaxing the greedy objective alone;
(iii)~uniform random selection of $k$ tasks;
(iv)~\emph{diversity-only} selection, which greedily maximizes $\mathrm{Div}(S)$
with no predictive stage; and
(v)~taxonomy-stratified random selection, which draws tasks at random across
taxonomy strata.
Randomized controls are averaged over three seeds; deterministic methods are run
once.
We run the controls on two datasets: the broader filtered logs with mean
imputation (\texttt{min\_completion\_rate}$=25\%$; 126 execution IDs and 852
  nonconstant candidate tasks after filtering, missing outcomes filled by
training-fold per-task means),\footnote{This export is counted by execution
  ID rather than by configuration and is therefore larger than the $n=67$
  sample of the principal sweep; the
  two analyses answer different questions, and we do not compare absolute $R^2$
across them.} and the complete-case matrix described next.

\smallskip
\paragraph{Complete-case common-reference matrix (RQ4)}
Because $y_c$ is computed over a configuration-specific task set, we also
built a complete execution-by-task matrix with \emph{no} missing cells:
99 execution IDs by 715 common tasks (70{,}785 observed outcomes), with 595
selection candidates after removing zero-variance tasks.
Every target and prediction is thus computed on the same observed reference
set, without imputation.
The 99 execution IDs correspond to only 46 distinct treatment definitions
(internally, \emph{flights}), the rest being repeated executions, so we present
this as a complete-case sensitivity analysis; it is not an independent replication.
On this matrix we run the $\epsilon$ sweep under all three validation
strategies at six task budgets, and all five controls at $k{=}10$.

\subsection{RQ1: How much predictive accuracy does selecting for taxonomic
coverage cost?}
\label{sec:results-reg}
Most directly, we fix a contemporaneous configuration family and ask: if we reduce the full benchmark to $k$ tasks, how much
predictive accuracy do we lose?
Figure~\ref{fig:final-r2-by-method} shows that the optimistic $R^2$---where
selection and scoring share one configuration family---stays high and falls only slightly as $\epsilon$ grows, while Figure~\ref{fig:div-by-method} shows the
taxonomy diversity score $\mathrm{Div}(S)$ rising steeply and saturating by
$\epsilon\approx0.1$--$0.2$.
Together, the two show that near-complete coverage of the task taxonomy costs
almost nothing in predictive accuracy over a known configuration
family.

The stricter question---how well the reduced benchmark predicts
\emph{new} configurations---is where the honest $R^2$ of
Figure~\ref{fig:r2-weighted-by-method} enters, and the absolute numbers drop
sharply: honest $R^2$ is well below its optimistic counterpart across the
sweep, and under the walk-forward strategies it is sometimes \emph{negative}
(Section~\ref{sec:results-hierarchy}).
We therefore interpret the honest curves \emph{relatively}, as a
regularization diagnostic rather than a claim of strong prediction on new
configurations. Within that interpretation they are informative: the honest
curve is frequently
\emph{non-monotone} in $\epsilon$, rising to a peak at intermediate $\epsilon$ before falling,
whereas the optimistic $R^2$ decreases almost monotonically.
This divergence is a regularization signature: pure greedy selection ($\epsilon{=}0$)
over-optimizes for training correlations that do not transfer, and a moderate
$\epsilon$ partly corrects it, so under out-of-fold evaluation $R^2$
maximization is \emph{not} the best selection rule.
Table~\ref{tab:headline-reg} quantifies the effect at $\epsilon{=}0$
vs.\ $\epsilon{=}0.1$ in four cells of the $k{=}10$ grid: in
three of four the diversity tie-breaker improves honest $R^2$, and in
the first it turns a negative $R^2$ positive---though several cells remain
negative in absolute terms, so the gain is only reduced overfitting.

\begin{table}[t]
  \centering
  \setlength{\tabcolsep}{2pt}
  \caption{Honest $R^2$ at $\epsilon{=}0$ vs.\ $\epsilon{=}0.1$, $k{=}10$,
    across four representative cells. Moderate $\epsilon$ can reduce
  selection-bias overfitting.}
  \label{tab:headline-reg}
  \begin{tabular}{p{2cm}p{0.8cm}ccp{3.5cm}}
    \toprule
    \textbf{Scenario} & \textbf{\texttt{min compl.}} &
    \textbf{$\epsilon{=}0$} & \textbf{$\epsilon{=}0.1$} & \textbf{$\Delta$
    honest $R^2$} \\
    \midrule
    Nested $K$-Fold          & 1\%  & $-0.157$ & $+0.059$ & $+0.22$
    (neg.\ $\to$ pos.) \\
    Nested $K$-Fold          & 5\%  & $+0.050$ & $+0.159$ & $+0.11$
    ($\approx\!3\times$) \\
    Walk-Forward Sliding     & 5\%  & $-1.242$ & $-0.277$ & $+0.97$
    ($4.5\times$ less negative) \\
    Walk-Forward Expanding   & 5\%  & $-0.253$ & $-0.316$ & $-0.06$ (slightly worse) \\
    \bottomrule
  \end{tabular}
\end{table}

Past the early band the two curves diverge: coverage saturates
(Figure~\ref{fig:div-by-method})---once the band reaches most taxonomy bins,
widening it adds little coverage---while honest $R^2$ continues to decline past
$\epsilon\approx0.2$--$0.3$.

A second measure supports this interpretation: the gap between optimistic and honest
$R^2$---a direct measure of selection-bias overfitting---is universally
positive, largest for the walk-forward methods and smallest for K-Fold, and
moderate $\epsilon$ tends to shrink it.

\subsection{RQ2: When does coverage help most?}
\label{sec:results-regime}
The regularization benefit is not uniform; it concentrates in the regimes where
greedy selection is most prone to overfitting.
Figures~\ref{fig:kmin-grid-kfold}--\ref{fig:kmin-grid-wfsliding2} in the appendix show the honest $R^2$ surface over the full $k \times$
\texttt{min\_completion\_rate} grid as a function of $\epsilon$, for each
validation strategy, and two patterns emerge.
First, the benefit is concentrated in the
\emph{low-\texttt{min\_completion\_rate} rows}---the runs in which we admit
configurations even though many of their task outcomes are missing.
Noisier configurations make the inner cross-validation scores noisier too, so
the tiny $R^2$ gaps that the pure greedy approach chases are mostly noise,
and breaking those
near-ties by taxonomic diversity helps more---the regime where the non-monotone
honest-$R^2$ curve is clearest (the low-\texttt{min\_completion\_rate} rows of
Table~\ref{tab:headline-reg}).
Second, the benefit is concentrated in the \emph{small-$k$ columns}: at
$k\in\{7,10\}$ the $\epsilon$ sweep shows visible gains, while at
$k\in\{20,30\}$ the curves flatten and the choice of $\epsilon$ has much less
effect, because the model then has enough features to fit well regardless of
which tasks were selected.
Because a modest reduced benchmark already captures most of the recoverable
signal, the open problem is how to spend a \emph{small} budget well, which is where
the diversity tie-breaker helps most.

\subsection{RQ3: Does the reduced benchmark survive temporal shift?}
\label{sec:results-hierarchy}
Only partly.
\emph{Future} configurations are a stricter test than
interchangeable draws from one population, and the three validation instruments
of Section~\ref{sec:eval-setup} separate the two interpretations.
Figure~\ref{fig:r2-weighted-by-method} shows the ordering Nested $K$-Fold $>$
Walk-Forward Expanding $>$ Walk-Forward Sliding holding across the $\epsilon$
sweep, with Sliding often turning negative.
A negative honest $R^2$ means the model predicts worse than the mean $y_c$; it
arises because $n$ is small, the matrix is sparse, Ridge is deliberately linear,
and the sliding window forces generalization across substantial temporal
shift.
The two gaps of Section~\ref{sec:eval-setup} decompose this difficulty: the
K-Fold/Expanding gap measures the benefit of temporal leakage, and the
Expanding/Sliding gap the cost of discarding older history.

\subsection{RQ4: Is the benefit due to the taxonomy, and does it survive a
common reference task set?}
\label{sec:results-controls}
Yes on both counts on the complete-case matrix, where every configuration is
scored on the same 715 tasks; on the broader, imputed logs the coverage gain
holds but the predictive advantage over random choice within the band does not
hold at every $\epsilon$.

\smallskip
\paragraph{Complete-case matrix}
The diversity--accuracy pattern of RQ1 reproduces on the zero-missing matrix
(Table~\ref{tab:complete-case-baselines},
App.~\ref{app:complete-case-controls}). At $k{=}10$, moving from
$\epsilon{=}0$ to $\epsilon{=}0.05$ raised the diversity of the reduced
benchmark selected on all data from $0.649$ to $0.815$ and improved honest
weighted-Ridge $R^2$ under all three validation strategies.
Across the six task budgets, moving from $\epsilon{=}0$
to $\epsilon{=}0.1$ increased diversity in all 18 validation-by-budget
comparisons and honest $R^2$
in 13 of the 18.
Since every cell is observed and every configuration is scored on the same 715
tasks, the pattern cannot come from imputation or from differences in which
tasks ran.
Among the controls, taxonomy-guided selection substantially increases
diversity relative to pure
greedy, uniform random, and $\epsilon$-band random selection at the same budget.
Diversity-only selection maximizes coverage but predicts poorly, so the
predictive stage is necessary.
And taxonomy-guided selection outperforms its matched $\epsilon$-band-random
control---same band, random choice within it---under all three validation
strategies at both $\epsilon{=}0.05$ and $\epsilon{=}0.1$, so the gain over
pure greedy selection comes from choosing within the band by taxonomic diversity.

\smallskip
\paragraph{Broader, imputed logs}
On the broader filtered logs, taxonomy guidance again substantially increases
diversity and usually improves honest $R^2$ over matched $\epsilon$-band random
selection: at $\epsilon{=}0.05$ it is higher under all three validation
strategies.
The one exception is Nested $K$-Fold at $\epsilon{=}0.1$, where $\epsilon$-band
random is marginally higher ($0.401$ vs.\ $0.396$).
The complete-case result is uniform and the imputed result is not, which
suggests that the taxonomy's advantage over random choice within the band
depends on $\epsilon$, on missingness, and on how well calibrated the inner-CV
predictor is; a systematic analysis is future work.
Our claim is therefore narrower: taxonomy guidance gives a controllable
increase in coverage at little or no cost in honest predictive
accuracy, and, on a fully observed common reference set, predicts better than
matched random choice within the same band.

Across the four questions, then, the evidence supports the finding anticipated
in Section~\ref{sec:test-selection}: the diversity tie-breaker behaves as a regularizer
over greedy selection rather than a trade-off---we discuss what this means for practice in Section~\ref{sec:discussion}.

\section{Discussion}
\label{sec:discussion}

\subsection{Interpreting the Results}

\paragraph{The diversity tie-breaker as a regularizer}
Across RQ1--RQ3, the diversity tie-breaker is better understood not as a
trade-off but as a \emph{regularizer over greedy selection}: pure greedy
selection ($\epsilon{=}0$) over-optimizes training correlations that do not transfer to
held-out configurations, and breaking near-ties by taxonomic diversity partly
corrects this.

RQ4 attributes the correction to the taxonomy: random choice within the same
band does not reproduce the coverage gain, and on the complete-case matrix it
predicts worse.
Ridge regularizes the learned weights, and the tie-breaker regularizes which
tasks greedy selection picks; both give up a little fit on the selection data
for better held-out prediction.

\paragraph{When a reduced benchmark can be trusted}
The honest-$R^2$ results limit \emph{what} a reduced benchmark should be used for.
For \emph{contemporaneous} configurations---ablations, hyperparameter
variants, release candidates of one base model---it is a sound instrument for
summarizing or ranking them, and a taxonomy-spanning one has construct and
face validity that a coverage-blind top-$k$ lacks, resists redundancy, and is
harder to game if the full-benchmark aggregate is later reweighted.
Across time, we advise caution: the low-rank structure the method assumes holds
under the independent and identically distributed (i.i.d.)\ assumption but
not stably as configurations evolve (the
Walk-Forward Sliding $R^2$ turns negative).

\paragraph{Task-count reduction}
The compression we study is steep in \emph{task count}: $k\in\{7,\dots,30\}$
tasks drawn from roughly a thousand (about $100\times$ at $k{=}10$), with the
saving recurring on \emph{every} evaluation while the curation cost is paid
once.
Task count is not cost, however: tasks differ substantially in runtime and
token usage, and the log snapshot we analyze does not include token-usage telemetry or comparable wall-clock measurements.
Our claims are therefore about task-count reduction, not measured monetary,
token, or runtime savings, which depend on the heterogeneous costs of the
selected tasks; a cost-aware selector (a budget in tokens or minutes rather
than tasks) and prospective measurement are future work.

\subsection{Implications}

\paragraph{For agent teams: choosing $\epsilon$ and refreshing}
A small tolerance band, $\epsilon$ of roughly $0.05$--$0.1$, is the operating
point we recommend.
At a moderate $\epsilon$ the reduced benchmark covers nearly the whole taxonomy
with no loss in honest $R^2$ relative to pure greedy selection, and often a small gain
that, as the absolute numbers make clear, reflects reduced overfitting rather
than strong prediction on new configurations; beyond
$\epsilon\approx0.2$--$0.3$ coverage has saturated while honest $R^2$ keeps
falling.
No single $\epsilon$ is uniformly optimal: $0.1$ worked best on the principal
sweep and $0.05$ was more reliable on the complete-case matrix, so the
regularizer's strength should be tuned to the noise level and budget at hand,
with the strongest case for moderate $\epsilon$ in the noisy,
low-\texttt{min\_completion\_rate}, small-$k$ regimes of
Section~\ref{sec:results-regime}.
Because the reduced benchmark does not transfer stably across time
(Section~\ref{sec:results-hierarchy}), it should also be
\emph{refreshed}---reselected---when the configuration population drifts,
rather than kept as a fixed proxy for future configurations.

\paragraph{For agent teams: weighting the taxonomy}
Our diversity metric weights every taxonomy dimension equally, but a team
shipping charting features cares more about visualization-related task types
and \emph{LLM capabilities} than about the full range of \emph{header type}s.
The metric extends to a relevance-weighted mean, with weights stated by the
team or derived from production telemetry (the task types users issue, and
especially those where the agent fails or users retry); the change reshapes
$\mathrm{Div}(\cdot)$ but leaves the predictive band $B_\epsilon$ untouched,
so the regularizer interpretation still holds.

\paragraph{For agent teams: auditing coverage gaps}
Apart from selection, the dimensions of Section~\ref{sec:taxonomy}
provide a coordinate system for \emph{auditing} an
evaluation portfolio: labeling both existing benchmark tasks and a sample of
production traffic along the same dimensions reveals regions of the task space
that users exercise but the benchmarks under-cover.
Teams at Microsoft used the taxonomy this way, applying it to
telemetry and internal benchmarks to identify task types and workbook
conditions their benchmarks did not adequately cover.
Gap analysis thus identifies \emph{which} regions need new tasks, while
taxonomy-aware selection compresses what is already covered.

\paragraph{For benchmark builders}
Our descriptive analysis (Section~\ref{sec:evaluation}) is itself actionable:
SpreadsheetBench, one of the most widely used public spreadsheet benchmarks,
is dominated by \emph{Formulas, Logic \& Automation} (a plurality the
rubric's catch-all tendency may inflate; Section~\ref{sec:evaluation}), almost all
\emph{Well-Specified}, and barely probes regions that are important in
practice---data cleaning over noisy workbooks, underspecified or ambiguous
instructions (Fig.~\ref{fig:sb-distributions}).
We encourage benchmark builders to populate these regions deliberately and to
report coverage along taxonomy dimensions rather than raw task count, so that size is not confused with breadth.

\paragraph{For other agent domains}
Our approach is not specific to spreadsheets: agent benchmarks in other
domains (coding, web, desktop) likewise define ad-hoc task categories without a
taxonomy designed for coverage-oriented selection (Section~\ref{sec:relwork}).
Practitioners there could follow the same construction
process~\citep{nickerson2013method}, label their tasks, and apply
taxonomy-aware minimization; validating this is future work.

\subsection{Limitations}
Several caveats temper these results, most rooted in the small, noisy logs.
\emph{Few, correlated configurations, one source.}
All three honest-$R^2$ estimators are high-variance (especially Walk-Forward
Sliding), and the rows of $X$ are not independent: several executions can
share one treatment (99 executions for 46 treatments in the complete-case
matrix), which weakens the i.i.d.\ assumption behind Nested $K$-Fold.
All evidence comes from one industrial source; replication on another agent
benchmark is needed before generalizing.
\emph{Missing data and the target.}
In the principal analysis each $y_c$ averages over its own configuration-specific
task set and missing cells are mean-imputed, which attenuates the coefficients of
heavily-missing columns and may not match the missingness mechanism.
The complete-case analysis (RQ4) removes both confounds within its rectangle,
but may not generalize to the excluded executions and tasks, and imputation can
still affect the exact estimates in the broader analysis.
\emph{Label reliability.}
The labels on which $\mathrm{Div}(S)$ depends are LLM-assigned and were
spot-checked by hand on a sample (Section~\ref{sec:evaluation}); we did not compute a formal agreement statistic, and a stratified two-rater study with
per-dimension agreement ($\kappa$) remains future work.
Because they are used only in the tie-break, label noise cannot change the
predictive band $B_\epsilon$.
The honest-$R^2$ results use task outcomes, not labels, so the benefit we
measured was obtained despite imperfect labels.
The coverage figures, computed from the same labels, are the one quantity
noise could overstate.
With uninformative labels the tie-breaker would degrade to $\epsilon$-band
random choice, so beating that control on the complete-case matrix is
indirect evidence that
the labels are informative.
A perturbation analysis of label errors would quantify how much noise the
tie-break tolerates.
\emph{The metric.}
$R^2$ is unstable when $\mathrm{Var}(y)$ is small, so mean absolute error or
rank correlation
may be more informative; we treat the validation hierarchy as a diagnostic.
A more reliable reduced benchmark needs more configurations, a
stable target (a fully observed reference set, as in RQ4), less selection
freedom (smaller $k$, stronger $\alpha$), $\epsilon$ tuned to the noise level,
and reporting which tasks recur across outer folds.

\section{Conclusion}

Our \emph{taxonomy of spreadsheet tasks} gives benchmark coverage a concrete,
externally observable meaning---the ``with respect to what'' that ad-hoc
benchmark categories lack---and our taxonomy-guided selector turns that meaning
into reduced benchmarks.
On real logs from Microsoft, selecting for taxonomic coverage costs little
predictive accuracy and often gains it, behaving as a \emph{regularizer} over
greedy selection, strongest where data is noisy and budgets are small;
matched-budget controls rule out a relaxed greedy objective as the source of the gain, and a complete-case replication rules out imputation.
Honest evaluation also marks the limits---poor transfer across a long time gap
when trained on little recent history, and a reduction measured in task count,
not yet in cost---so the practical recipe is a small,
taxonomy-spanning reduced benchmark, refreshed as the configurations drift.

\subsubsection*{\textbf{Data Availability}}
\label{sec:data-availability}

\noindent The benchmark logs are confidential and cannot be shared.
We will release the analysis code, with confidential identifiers removed,
an input schema, run commands, seeds, hyperparameters, and instructions for
rerunning it on another configuration-by-task matrix.

\subsubsection*{\textbf{Acknowledgments}}

\noindent The authors used AI assistants (GitHub Copilot and Claude Code, on
Anthropic's Claude and OpenAI's GPT models) to draft portions of the manuscript
from our notes and analyses, to write code behind many of the numbers and
figures (which we checked), and to copy-edit prose
under our direction; LLMs are also a component of the taxonomy-construction
method (Section~\ref{ssec:methodology}).
All research questions, study design, analysis decisions, and interpretations
are our own, and we take full responsibility for the paper's content.

\bibliographystyle{IEEEtranN}
\bibliography{references-short}

\clearpage
\appendices

\section{Complete Taxonomy Value Sets}
\label{app:taxonomy-values}

Table~\ref{tab:taxonomy} lists only \emph{example} measures and values for each dimension. For completeness, we enumerate below the full value set of every close-ended (categorical or multi-label) measure \emph{that appears in Table~\ref{tab:taxonomy}}, grouped by the three tiers of the taxonomy. Numeric measures (e.g., table and workbook size, or task difficulty), which take ranges rather than discrete values, and open-ended measures (e.g., data types, language, and free-text operation phrases), which admit no fixed value set, are therefore not listed. Where a per-value definition is given, it is taken (lightly edited) from the labeling prompt or capability schema used to assign the value; bare enumerations are listed without definitions. These value sets are produced programmatically from the labeling configuration so that they stay in sync with the implementation.

\subsection{Table Properties}

\paragraph{Header type (2 values).} The structure of a table's header (see also Fig.~\ref{fig:spreadsheet-examples}).
\begin{itemize}[leftmargin=1.4em,itemsep=1pt,topsep=2pt]
  \item \textbf{Simple}: A single header row only.
  \item \textbf{Complex}: Multiple header rows or one or more header columns present.
\end{itemize}

\paragraph{Structure (4 values).} The layout of a table, as detected from its orientation and headers.
\begin{itemize}[leftmargin=1.4em,itemsep=1pt,topsep=2pt]
  \item \textbf{columnMajorFlat}: Flat table with headers in the first row and data in columns below.
  \item \textbf{rowMajorFlat}: Flat table with headers in the first column and data in rows to the right.
  \item \textbf{hierarchical}: Table with nested or multi-level row or column headers.
  \item \textbf{other}: A table structure that does not fit the standard patterns.
\end{itemize}

\subsection{Spreadsheet and Workbook Properties}

\paragraph{Workbook primary intent (31 values).} The purpose of the workbook (the \emph{Domain/Topic} dimension), assigned by an LLM from the workbook content.
\begin{itemize}[leftmargin=1.4em,itemsep=1pt,topsep=2pt]
  \item \textbf{Financial Management}: budget, revenue, expenses, financial reporting, cash flow, profit and loss, balance sheet, financial planning, investment, funding sources
  \item \textbf{Project Planning}: project management, timeline, task tracking, milestone, resource allocation, Gantt chart, progress monitoring, project schedule
  \item \textbf{Data Analysis}: data analysis, statistical analysis, data visualization, summary statistics, pivot table, data aggregation, trend analysis, comparative analysis
  \item \textbf{Regulatory Compliance}: compliance, regulatory reporting, audit, legal compliance, certification, documentation, validation, reporting requirements
  \item \textbf{Human Resources}: payroll, salary, employee information, attendance, leave management, compensation, benefits, staff, personnel costs
  \item \textbf{Inventory Tracking}: inventory, stock management, product catalog, item code, quantity, procurement, supply chain, warehouse
  \item \textbf{Academic Administration}: student data, academic year, enrollment, grade, course, subject, teacher, school, educational assessment
  \item \textbf{Healthcare Records}: patient data, diagnosis, procedure code, healthcare provider, medical billing, insurance, clinical data, treatment, hospital
  \item \textbf{Event Scheduling}: calendar, schedule, event, dates, planning, timetable, session, participant, attendance
  \item \textbf{Construction Management}: construction, project cost, bill of quantities, contractor, materials, site location, infrastructure, building type
  \item \textbf{Grant and Funding}: grant application, funding request, grant allocation, award, program funding, grant reporting, subrecipient
  \item \textbf{Risk Assessment}: risk management, risk assessment, risk register, mitigation, failure analysis, safety, audit, action plan
  \item \textbf{Sales and Marketing}: sales, product, customer, order, pricing, marketing, campaign, revenue, sales tracking
  \item \textbf{Community Engagement}: community development, stakeholder, outreach, engagement, social impact, volunteerism, public health
  \item \textbf{Sustainability Reporting}: sustainability, environmental impact, energy consumption, waste management, greenhouse gas emissions, social responsibility
  \item \textbf{Transportation Data}: transportation, transit operations, vehicle types, route, ridership, infrastructure, maintenance, public transport
  \item \textbf{Scientific Research}: experimental data, gene expression, sample identification, biological data, measurement, data analysis, research study
  \item \textbf{Legal and Contractual}: contract, agreement, legal status, contract type, terms, compliance, documentation, approval
  \item \textbf{Demographic Analysis}: population, age group, gender, region, demographic data, census, household, survey
  \item \textbf{Insurance Administration}: insurance, premium, coverage, claim, policy, benefit, deductible, plan, provider
  \item \textbf{Real Estate Management}: property, address, value, rent, ownership, real estate, property type, appraisal
  \item \textbf{Resource Allocation}: resource planning, resource allocation, bandwidth, man-hours, team, assignment, utilization
  \item \textbf{Educational Resources}: textbook, curriculum, lesson, teaching materials, resource type, subject area, academic module
  \item \textbf{Data Entry and Validation}: data entry, validation, checklist, template, instructions, error checking, data quality
  \item \textbf{Sports Management}: team, match, league, player, score, tournament, competition, performance tracking
  \item \textbf{Procurement and Purchasing}: procurement, purchasing, vendor, supplier, order, invoice, contract, product catalog
  \item \textbf{Taxation and Accounting}: tax, income tax, deduction, accounting, journal entry, financial statement, tax calculation
  \item \textbf{Healthcare Services}: healthcare, service type, facility, provider, patient, treatment, care, reimbursement
  \item \textbf{Organizational Directory}: organization, contact information, address, phone number, email, staff directory, department
  \item \textbf{Metadata and Structure}: worksheet structure, metadata, creator information, modification date, data structure, spreadsheet type
  \item \textbf{Categorization and Taxonomy}: taxonomy, classification, category, code mapping, data categorization, thematic analysis, spreadsheet taxonomy
\end{itemize}

\paragraph{NAICS industry sector (20 values).} A complementary \emph{Domain/Topic} measure: the 2-digit North American Industry Classification System sector inferred from the workbook content.
\begin{itemize}[leftmargin=1.4em,itemsep=1pt,topsep=2pt]
  \item \textbf{11 -- Agriculture, Forestry, Fishing and Hunting}
  \item \textbf{21 -- Mining, Quarrying, and Oil and Gas Extraction}
  \item \textbf{22 -- Utilities}
  \item \textbf{23 -- Construction}
  \item \textbf{31 -- Manufacturing}
  \item \textbf{42 -- Wholesale Trade}
  \item \textbf{44 -- Retail Trade}
  \item \textbf{48 -- Transportation and Warehousing}
  \item \textbf{51 -- Information}
  \item \textbf{52 -- Finance and Insurance}
  \item \textbf{53 -- Real Estate and Rental and Leasing}
  \item \textbf{54 -- Professional, Scientific, and Technical Services}
  \item \textbf{55 -- Management of Companies and Enterprises}
  \item \textbf{56 -- Administrative and Support and Waste Management and Remediation Services}
  \item \textbf{61 -- Educational Services}
  \item \textbf{62 -- Health Care and Social Assistance}
  \item \textbf{71 -- Arts, Entertainment, and Recreation}
  \item \textbf{72 -- Accommodation and Food Services}
  \item \textbf{81 -- Other Services (except Public Administration)}
  \item \textbf{92 -- Public Administration}
\end{itemize}

\subsection{Task and Instruction Properties}

\paragraph{Instruction specificity (5-point scale).} The \emph{Clarity} dimension, rated against a fixed rubric.
\begin{itemize}[leftmargin=1.4em,itemsep=1pt,topsep=2pt]
  \item \textbf{1 (Very Ambiguous)}: Impossible to act on without significant clarification. Lacks any specific target or action.
  \item \textbf{2 (Ambiguous)}: Missing critical information. Has multiple plausible but different interpretations.
  \item \textbf{3 (Reasonably Specified)}: The main intent is clear, but minor assumptions are needed.
  \item \textbf{4 (Well-Specified)}: The query is clear, direct, and contains enough information to be executed with high confidence.
  \item \textbf{5 (Very Well-Specified)}: Perfectly clear and unambiguous, likely referencing specific columns, sheets, or values, leaving no room for interpretation.
\end{itemize}

\paragraph{Meta operation -- operation types (35 values).} The action component of the \emph{Operations} dimension; a task may carry several.
\textit{add}, \textit{align}, \textit{allocate}, \textit{calculate}, \textit{compare}, \textit{copy}, \textit{count}, \textit{create}, \textit{disable}, \textit{display}, \textit{fill}, \textit{filter}, \textit{fix}, \textit{format}, \textit{freeze}, \textit{hide}, \textit{identify}, \textit{link}, \textit{merge}, \textit{navigate}, \textit{paste}, \textit{protect}, \textit{refresh}, \textit{remove}, \textit{rename}, \textit{replace}, \textit{resize}, \textit{retain}, \textit{save}, \textit{sort}, \textit{split}, \textit{summarize}, \textit{unmerge}, \textit{validate}, \textit{None}.

\paragraph{Meta operation -- operation objects (44 values).} The object each operation acts upon; a task may carry several.
\textit{aggregation}, \textit{alignment}, \textit{border}, \textit{cell}, \textit{cell reference}, \textit{chart}, \textit{column}, \textit{comment}, \textit{connection}, \textit{dashboard}, \textit{data}, \textit{error}, \textit{filter}, \textit{footer}, \textit{format}, \textit{formula}, \textit{header}, \textit{hyperlink}, \textit{image}, \textit{macro}, \textit{Machine Learning Model}, \textit{named range}, \textit{number}, \textit{object}, \textit{page layout}, \textit{pane}, \textit{picture}, \textit{PivotTable}, \textit{protection}, \textit{row}, \textit{shape}, \textit{size}, \textit{slicer}, \textit{sparkline}, \textit{style}, \textit{table}, \textit{text box}, \textit{theme}, \textit{tool}, \textit{validation rule}, \textit{value}, \textit{workbook}, \textit{worksheet}, \textit{None}.

\paragraph{Task type -- action type (3 values).} The coarsest operationalization of \emph{Task type}: whether the instruction asks for information or directs an action.
\begin{itemize}[leftmargin=1.4em,itemsep=1pt,topsep=2pt]
  \item \textbf{ask}: Questions about insights, explanations, or recommendations without an explicit instruction to modify the workbook.
  \item \textbf{act}: Directives to create, edit, format, calculate, or otherwise change or produce spreadsheet artefacts.
  \item \textbf{unknown}: The request could be interpreted either way or lacks sufficient detail.
\end{itemize}

\paragraph{Task type -- intent themes (12 values).} One operationalization of \emph{Task type}, derived by thematic analysis of the instruction text.
\begin{itemize}[leftmargin=1.4em,itemsep=1pt,topsep=2pt]
  \item \textbf{Filtering \& Selection}: Filtering, subsetting, excluding, or deduplicating records/entities based on criteria or thresholds.
  \item \textbf{Aggregation \& Summarization}: Summing, counting, frequency, totals, breakdowns, subtotals, and high-level numeric summaries.
  \item \textbf{Grouping \& Categorization}: Grouping, segmenting, classifying, labeling categories or entities for organization or comparative frames.
  \item \textbf{Comparison, Ranking \& Extremes}: Comparing entities/groups/time periods, ranking/order, identifying max/min/top/bottom/extremes and differences.
  \item \textbf{Statistics, Percentages \& Metrics}: Averages, medians, ratios, percentages, rates, statistical and correlation measures.
  \item \textbf{Temporal \& Trend Analysis}: Time-based or sequential analysis: trends, patterns over periods, period-over-period comparisons.
  \item \textbf{Extraction, Lookup, Join \& Reference}: Retrieving specific values, lookups, joins, merges, matching, referencing or attributing data.
  \item \textbf{Data Cleaning, Quality \& Error Handling}: Handling missing/invalid data, validation, error/outlier/anomaly detection, deduplication for quality, formula error resolution.
  \item \textbf{Transformation, Feature \& Text Engineering}: Transforming structure or formats, feature engineering, mapping, text parsing/splitting/concatenation, numeric/date conversions, sequence generation.
  \item \textbf{Visualization, Formatting \& Presentation}: Charts, graphs, dashboards, conditional formatting, highlighting, styling, and presentational layout.
  \item \textbf{Formulas, Logic \& Automation}: Creating/modifying formulas, conditional logic, automation (macros/VBA), dynamic calculation behavior.
  \item \textbf{Structure, Layout \& Data Organization}: Managing tables, pivot tables, sheets, columns, layout, data movement/placement, consolidation, access/permissions.
\end{itemize}

\paragraph{LLM capabilities (8 categories, 20 subcategories).} The agent skills a task requires; also serves as the capability-based operationalization of \emph{Task type}. Listed as category (with its subcategories).
\begin{itemize}[leftmargin=1.4em,itemsep=2pt,topsep=2pt]
  \item \textbf{Analysis} --- All forms of data analysis and exploration
  \begin{itemize}[leftmargin=1.2em,itemsep=1pt,topsep=1pt]
    \item \textit{Analyze Descriptively}: Basic Multi-Dimensional Analysis (MDA) operations on tabular entities
    \item \textit{Analyze Advanced}: Sophisticated analytical techniques determined by schema and relationships
  \end{itemize}
  \item \textbf{Visualization} --- Creating visual representations and interactive data exploration tools
  \begin{itemize}[leftmargin=1.2em,itemsep=1pt,topsep=1pt]
    \item \textit{Create Charts}: Creating, modifying, or analyzing charts and visualizations
    \item \textit{Create PivotTables}: Creating, modifying, or analyzing PivotTables and PivotCharts
  \end{itemize}
  \item \textbf{Data Management} --- Operations for acquiring, organizing, and maintaining data
  \begin{itemize}[leftmargin=1.2em,itemsep=1pt,topsep=1pt]
    \item \textit{Import Data}: Importing data from files that Excel handles natively
    \item \textit{Clean Data}: Cleaning, organizing, or restructuring messy data
    \item \textit{Shape Data}: Transforming data structure and layout
    \item \textit{Create Tables}: Creating structured tables or converting ranges to tables
    \item \textit{Validate Data}: Implementing data validation rules or checking data quality
    \item \textit{Style Data}: Applying visual formatting and styling to data and worksheets
  \end{itemize}
  \item \textbf{Computation} --- Mathematical operations and calculations
  \begin{itemize}[leftmargin=1.2em,itemsep=1pt,topsep=1pt]
    \item \textit{Use Formulas}: Creating, modifying, or using Excel formulas and functions
    \item \textit{Run Python}: Using Python within Excel for advanced computations
  \end{itemize}
  \item \textbf{Content Processing} --- Working with text and non-tabular content
  \begin{itemize}[leftmargin=1.2em,itemsep=1pt,topsep=1pt]
    \item \textit{Analyze Text}: Analyzing, processing, or manipulating text data
    \item \textit{Summarize Text}: Summarizing or condensing text content
    \item \textit{Translate Text}: Translating text between different languages
    \item \textit{Process Images}: Working with images, PDFs, or visual content
  \end{itemize}
  \item \textbf{Integration} --- Connecting with external systems and collaboration
  \begin{itemize}[leftmargin=1.2em,itemsep=1pt,topsep=1pt]
    \item \textit{Use M365}: Using Microsoft 365 integrations and cloud services
    \item \textit{Connect External}: Connecting to external systems and data sources
    \item \textit{Collaborate}: Adding comments, sharing, or collaborative features
  \end{itemize}
  \item \textbf{Automation} --- Automating repetitive tasks and workflows
  \begin{itemize}[leftmargin=1.2em,itemsep=1pt,topsep=1pt]
    \item \textit{Automate Tasks}: Creating macros, VBA scripts, or other automation
  \end{itemize}
  \item \textbf{Other} --- Actions that don't fit into other specific categories
\end{itemize}

\section{Toy Reduced-Benchmark Example}
\label{app:toy-example}

Suppose a benchmark of 1{,}000 spreadsheet tasks, with configurations varying
primarily along two latent capabilities: \emph{formula reasoning} and
\emph{data cleaning}.
A configuration strong on formula reasoning but weak on data cleaning passes
most formula tasks (say, 600/700) but few data-cleaning tasks (50/300), for an
overall pass rate near 0.65; a configuration with the opposite profile reaches a
similar rate for opposite reasons.
Because pass rate is an average over outcomes, it can in principle be
reconstructed from any small set of tasks that captures both factors---here, two
well-chosen tasks plus a learned weighting from their outcomes to the
full-benchmark score.
Conversely, if 200 formula tasks measure the same sub-skill, dropping 199 of
them costs almost nothing in predictive accuracy; the survivor represents its
cluster.

Table~\ref{tab:toy} makes this concrete on a stylized $4\times6$ example in
which F1--F2 and D1--D3 are perfectly correlated: four of the six tasks are
redundant, and the two-task reduced benchmark $\{F1,D1\}$ reconstructs every
configuration's full-benchmark pass rate exactly.

\begin{table}[t]
\centering
\setlength{\tabcolsep}{2pt}
\caption{Stylized $4\times6$ instance of the mechanism described in the text: formula tasks F1--F3 and data-cleaning tasks D1--D3, full-benchmark pass rate $y_c$, and its estimate $\hat{y}_c$ from the two-task reduced benchmark $\{F1,D1\}$ with learned Ridge weights $\hat{\beta}=(\tfrac{1}{3},\tfrac{2}{3})$.}
\label{tab:toy}
\begin{tabular}{lcccccccc}
\toprule
& \multicolumn{3}{c}{\textbf{Formula tasks}} & \multicolumn{3}{c}{\textbf{Data-clean.\ tasks}} & \textbf{$y_c$} & \textbf{$\hat{y}_c$} \\
\cmidrule(lr){2-4} \cmidrule(lr){5-7}
\textbf{Config} & F1 & F2 & F3 & D1 & D2 & D3 & (full) & ($\{F1,D1\}$) \\
\midrule
$c_1$ (F$+$, D$-$) & 1 & 1 & 0 & 0 & 0 & 0 & 0.33 & 0.33 \\
$c_2$ (F$-$, D$+$) & 0 & 0 & 1 & 1 & 1 & 1 & 0.67 & 0.67 \\
$c_3$ (F$+$, D$+$) & 1 & 1 & 1 & 1 & 1 & 1 & 1.00 & 1.00 \\
$c_4$ (F$-$, D$-$) & 0 & 0 & 0 & 0 & 0 & 0 & 0.00 & 0.00 \\
\bottomrule
\end{tabular}
\end{table}

\section{Taxonomy Construction Details}
\label{app:taxonomy-method-details}

\paragraph{Choosing concrete measures}
With a dimension established, we must decide \emph{how to measure it}: which one
or more \emph{concrete measures} assign a task a value along the dimension, and
which values each measure allows.
We deliberately allow a single dimension to have more than one measure, because an axis of variation can often be operationalized in different ways
with distinct tradeoffs (the \emph{Task type} dimension below is our clearest
example).
This extends the single-characteristic-set view of
\citet{nickerson2013method}---in whose terms a measure's value set is the
dimension's set of \emph{characteristics}---and is reflected in the rightmost
column of Table~\ref{tab:taxonomy}.

\emph{Objective (structural) measures} can be extracted programmatically from
workbook metadata (e.g., cell counts, merged-cell detection, hierarchical
structure), without human judgment.
\emph{Subjective (interpretive) measures} require interpreting the task
instruction or spreadsheet context, and we operationalized them in one of two
ways depending on whether the measure's value set was emergent or fixed in
advance.
When the value set had to be \textbf{discovered} from the data, we used
LLM-assisted thematic analysis following \citet{naeem2025thematic}, who
validated their approach against manual thematic analysis.
For each task, we gave GPT-4.1 the axis to code along (e.g., task type), the
task instruction, and a sample of the workbook data, and the LLM produced
initial \textit{codes} for that task.
After coding all corpus tasks this way, we had the LLM group the per-task codes
into higher-level \textit{themes}, which became the measure's value set.
We reviewed these groupings and iterated on the prompts and codebook (e.g.,
after noticing a few instances of prompt bleeding/context leakage) until the
categories stabilized.
When the value set was \textbf{fixed in advance}, we instead had the LLM rate
each task against a predefined rubric---for instance, scoring instruction
specificity on a 1--5 scale---so no theming step was required.

The \emph{task type} dimension is our clearest case of a dimension admitting
several candidate measures.
We explored alternatives because it is arguably the most consequential for the
downstream evaluation, and because ``task type'' is not one obvious
concept---different operationalizations are study-design alternatives with
distinct tradeoffs.
We compared four, mirroring the deductive/inductive alternation above.
First, we applied bottom-up thematic analysis~\citep{naeem2025thematic} of the
natural-language instructions, producing fine-grained action-oriented categories
(e.g., Aggregation~\&~Summarization, Comparison~\&~Ranking,
Filtering~\&~Selection, Formulas~\&~Automation); these are well-grounded in the
data but may yield more categories than are practical for test selection.
Second, we tried LLM-amplified few-shot clustering~\citep{viswanathan2024large},
which is query-efficient---the LLM labels most items and routes uncertain cases
near the decision boundary to a human judge---but in our setting it yielded
coarser, more function-oriented groups (e.g., Data Analysis, Data Filtering,
Formula, Pivot Table) that risk merging meaningfully different tasks.
Third, product managers at Microsoft did top-down coding, defining task
types by the capabilities the agent requires.
This produced an agent-centric hierarchy (e.g., Descriptive Analysis,
Computation--Using Formulas, Visualization--Creating PivotTables) that is
directly useful for capability-gap analysis but reflects the product team's
framing and may not generalize beyond a specific agent.
Finally, we tried coding the workbook's primary intent rather than the
natural-language instruction, which produced purpose-oriented types (e.g.,
Data~Entry~\&~Validation, Financial~Management) but can conflate tasks whose
instructions differ substantially yet operate on similar workbooks.
Thus, even for a single dimension, one can vary both \emph{what} is coded (the
natural language task instruction, the implied workbook intent, or the required
agent capabilities) and \emph{how} (thematic analysis, few-shot clustering, or
top-down decomposition), and each choice has different tradeoffs in
granularity, generalizability, and practical utility.

We faced a similar choice for \emph{Domain/Topic}.
Coding workbook intent, as above, groups tasks by what the spreadsheet is \emph{for}, but two tasks with the same purpose (e.g., financial
reporting) may involve very different data domains.
We therefore added a complementary operationalization that assigns, via LLM
labeling of the workbook content, a sector from the
\emph{North American Industry Classification System}
(NAICS),\footnote{\url{https://www.census.gov/naics/}} the standard
U.S.\ federal taxonomy for classifying businesses by industry (e.g., sector~52
\emph{Finance and Insurance}, sector~44--45 \emph{Retail Trade}).
NAICS sector provides a standardized, reproducible domain label, while workbook
intent captures purpose-level variation that sector codes miss; keeping both
gives the taxonomy finer-grained coverage of the Domain/Topic space.

\paragraph{Validation and termination}
Because the process is iterative, we needed criteria for when to stop.
We terminated refinement when the taxonomy satisfied two groups of ending
conditions~\citep{nickerson2013method}.
\emph{Objectively}: (i)~for single-label measures each task receives exactly one
value (e.g., header type is either \emph{simple} or \emph{hierarchical}, not
both); (ii)~for multi-label measures a task may receive several values, because
a single instruction can involve multiple operations or skills (e.g., a task's
meta operations may include \emph{lookup}, \emph{calculate}, and
\emph{visualize}); and (iii)~every task in the corpus is classified by every
dimension.
\emph{Subjectively}, we required the taxonomy to be concise, robust, comprehensive for
our scope, extendible, and explanatory for error
analysis~\citep{nickerson2013method}.
To check these conditions, two researcher co-authors reviewed the full
taxonomy, discussed disagreements, merged redundant categories, and clarified
definitions, incorporating practitioner feedback from internal presentations and
sessions with researchers and product managers at Microsoft to prioritize
evaluation-relevant distinctions~\citep{lupetti2024making}.

\section{Taxonomy Dimension Examples}
\label{app:taxonomy-examples}

\begin{figure}[t]
  \centering
  \scriptsize
  \subfigure[Simple header, flat layout.]{%
  \setlength{\tabcolsep}{5pt}
    \begin{tabular}[t]{|l|r|l|r|}
      \hline
      \textbf{Name} & \textbf{Age} & \textbf{City} & \textbf{Salary} \\
      \hline
      Alice & 30 & NYC & 75\,000 \\
      Bob   & 25 & LA  & 62\,000 \\
      Carol & 35 & Chicago & 88\,000 \\
      \hline
    \end{tabular}%
    \label{fig:ex-simple}%
  }
  \hfill
  \subfigure[Flat row-major layout.]{%
  \setlength{\tabcolsep}{5pt}
    \begin{tabular}[t]{|l|rrr|}
      \hline
      \textbf{Metric} & \textbf{Jan} & \textbf{Feb} & \textbf{Mar} \\
      \hline
      Revenue & 50 & 55 & 60 \\
      Cost    & 30 & 32 & 35 \\
      Profit  & 20 & 23 & 25 \\
      \hline
    \end{tabular}%
    \label{fig:ex-rowmajor}%
  }

  \subfigure[Hierarchical header, non-flat layout.]{%
  \setlength{\tabcolsep}{5pt}
    \makebox[0.5\columnwidth][c]{%
    \begin{tabular}[t]{|p{0.8cm}|rr|rr|}
      \hline
       & \multicolumn{2}{c|}{\textbf{2024 Q1}} & \multicolumn{2}{c|}{\textbf{2024 Q2}} \\
      \textbf{Region} & \textbf{Rev} & \textbf{Cost} & \textbf{Rev} & \textbf{Cost} \\
      \hline
      East & 120 & 80 & 135 & 85 \\
      West &  95 & 70 & 110 & 75 \\
      \hline
    \end{tabular}%
    }%
    \label{fig:ex-complex}%
  }
  \hfill
  \subfigure[Two table-like regions on one sheet.]{%
  \setlength{\tabcolsep}{5pt}
    \makebox[0.35\columnwidth][c]{%
    \begin{tabular}[t]{|p{0.4cm}|l|}
      \hline
      \textbf{ID} & \textbf{Item} \\
      \hline
      1 & Bolts \\
      2 & Nuts  \\
      \hline
    \end{tabular}%
    \quad
    \begin{tabular}[t]{|p{0.4cm}|r|}
      \hline
      \textbf{Cat} & \textbf{Total} \\
      \hline
      A & 340 \\
      B & 215 \\
      \hline
    \end{tabular}%
    }%
    \label{fig:ex-regions}%
  }
  \caption{Illustrative spreadsheet structures: (a)~flat column-major layout with a simple header; (b)~flat row-major layout; (c)~non-flat layout with a hierarchical header; (d)~two distinct table-like regions on one sheet.}
  \label{fig:spreadsheet-examples}
\end{figure}

The examples below use SpreadsheetBench~\citep{ma2024spreadsheetbench}, which is
public.
We refer to tasks by their SpreadsheetBench identifiers (e.g., \texttt{13-1}).
Because the original SpreadsheetBench tasks are phrased as how-to questions to
a chatbot, we rephrased them into imperative instructions for an AI agent while
keeping all other task details unchanged.

\subsection{Table properties}
The unit of analysis is the workbook, but several dimensions describe the
structure of individual tables within it; our SpreadsheetBench annotations
record these signals via per-table metadata and sheet-/workbook-level
aggregates.
The simpler structural measures---\emph{table size} (rows and columns),
\emph{vocabulary size} (number of unique values), \emph{data types} (e.g., date,
numeric, text), and \emph{language} (almost always English)---are as defined in
Table~\ref{tab:taxonomy}; two dimensions warrant illustration.
\emph{Header type} captures whether headers are simple
(Fig.~\ref{fig:spreadsheet-examples}(a)) vs.\ hierarchical with merged cells
(Fig.~\ref{fig:spreadsheet-examples}(c)).
\emph{Structure} captures table layout: \emph{flat column-major} means each
column is an attribute and each row a record
(Fig.~\ref{fig:spreadsheet-examples}(a)); \emph{flat row-major} transposes this
so that each row is an attribute and each column a record
(Fig.~\ref{fig:spreadsheet-examples}(b)); \emph{non-flat} covers crosstabs,
pivoted layouts, and other arrangements where neither reading direction alone
suffices (Fig.~\ref{fig:spreadsheet-examples}(c)).
Finally, \emph{data cleanliness} captures noise such as duplication or
inconsistent
formatting.\footnote{Data cleanliness is difficult to operationalize reliably and was excluded from our empirical analyses in Section~\ref{sec:evaluation}.}

\subsection{Spreadsheet and workbook properties}
\emph{Size} captures the scale of the parent spreadsheet or workbook as a
whole---its number of sheets, its number of table-like regions, and its on-disk
file size.
For example, task~\texttt{13-1} explicitly spans two sheets (``Combine data from
the `RANGES' sheet into the `LISTS' sheet \dots'') and its workbook contains six
table regions (three on each sheet), whereas task~\texttt{19-7} uses a single
sheet with a single table region (Fig.~\ref{fig:spreadsheet-examples}(d)
illustrates a sheet with multiple regions).
We include file size to capture scaling effects for the evaluation harness.
\emph{Structural complexity} aggregates these structural signals (e.g., size and
layout) into an overall rating.
For instance, a workbook with 5208 rows and five columns (task~\texttt{19-7})
likely represents a very different workload than one with 85 rows and nine
columns (task~\texttt{13-1}).
\emph{Domain/Topic} captures the subject matter and the primary intent of the
workbook (recall Section~\ref{ssec:methodology}).
For example, the workbook for task~\texttt{17-35} is a supplier invoice register
whose columns record dates, supplier names, tax-identification numbers, invoice
numbers, and monetary amounts for pharmaceutical vendors---placing it in NAICS
sector~42 \emph{Wholesale Trade}; its inferred workbook intent is
\textit{Data Filtering}.

\subsection{Task properties}
\emph{Clarity} captures how well-specified the instruction is, on a scale from 1
to 5: some tasks provide detailed ranges and constraints (e.g., clarity score~5
for task~\texttt{19-7}: ``split data from the `LASTROW' sheet into two separate
sheets named `MINUS' and `PLUS', based on negative and positive values found in
column E''), while others are underspecified.

\emph{Operations} capture the action--object intent signals in the instruction
at two levels.
At the surface level, \emph{query verb--noun phrases} record specific actions
and their objects, e.g., task~\texttt{13-1} includes phrases like ``delete old
ranges'' and ``add TOTAL row to sum amounts,'' while task~\texttt{19-7} includes
``create macro'' and ``delete existing sheets.''
At a more abstract level, \emph{meta operations} capture composable operation
primitives: task~\texttt{17-35} includes \texttt{filter} (data),
\texttt{validate} (cell), and \texttt{add} (column); task~\texttt{13-1} includes
\texttt{remove} (data), \texttt{merge} (data), and \texttt{add} (row); and
task~\texttt{19-7} includes \texttt{remove} (worksheet), \texttt{split} (data),
and \texttt{allocate} (cell).

\emph{Task type} captures the coarse-grained kind of work a task demands.
SpreadsheetBench distinguishes explicit questions (\texttt{ask}, e.g., ``Explain
what is wrong with the VBA code \dots'' in task~\texttt{209-30}) from action
requests (\texttt{act}, e.g., ``Create a macro to split data \dots'' in
task~\texttt{19-7}), and from cases where the instruction is too vague to
determine intent (\texttt{unknown} in task~\texttt{55224}).
\emph{Task difficulty} reflects both the task's horizon (the number of
operations/reasoning steps required) and its operation mix; for example,
task~\texttt{19-7} explicitly bundles deleting sheets, splitting data, placing
outputs into specific ranges, and inserting total rows---multi-step macro
creation over a \texttt{5208$\times$ 5} table---versus smaller single-formula
tasks.

\emph{LLM capabilities} capture the types of agent skills required, such as
automation and formula use (e.g., \texttt{Automation.Automate Tasks} and
\texttt{Computation.Use Formulas} in task~\texttt{19-7}).
Although some overlap exists between Task type and LLM capabilities (e.g., a
``data visualization'' task type often requires a ``charting'' capability), they
capture different facets: \emph{Task type} is grounded in properties of the
workbook--instruction input---consistent with our criterion of externally
observable properties---whereas \emph{LLM capabilities} reflects the
agent-centric view originating from Microsoft's product managers.
Keeping them apart lets the taxonomy distinguish \emph{what the user asks for}
from \emph{what the agent must be able to do}.

\section{Formalization of the Selection Task}
\label{app:formalization}

Let $\mathcal{C}$ be a set of configurations (e.g., candidate model variants).
Let $\mathcal{T}$ be the set of tasks.
For each configuration $c\in\mathcal{C}$ and task $t\in\mathcal{T}$ we observe
an outcome $r_{c,t}\in\{0,1\}$ (fail/pass), with missingness allowed when a task
was not executed.

We define the full-benchmark score for configuration $c$ as the mean success
across all tasks observed for that configuration:
\begin{equation}
 y_c \;=\; \frac{1}{|\mathcal{T}_c|}\sum_{t\in\mathcal{T}_c} r_{c,t},
\end{equation}
where $\mathcal{T}_c\subseteq\mathcal{T}$ denotes tasks with non-missing
outcomes for $c$.
Note that $\mathcal{T}_c$ is configuration-specific; the complete-case
analysis of Section~\ref{sec:results-controls} instead scores every configuration on
one common, fully observed $\mathcal{T}$.

\textbf{Goal.}
Choose a small subset $S\subset\mathcal{T}$, $|S|=k$, such that a score computed
from only tasks in $S$ accurately predicts $y_c$ across configurations.

\paragraph{Scope of $\mathcal{C}$ (which configurations do we care about?).}
A reasonable concern is that some configurations in historical logs are ``old''
(e.g., obsolete model generations) and may never be deployed again.
We do not assume that \emph{all} historical configurations are
decision-relevant.
Instead, $\mathcal{C}$ should be understood as the set of configurations
relevant to a particular evaluation goal (e.g., a current model family, a
specific tool stack, or a time window).
The predictive objective is then: \emph{within that scope}, can a small reduced
benchmark predict the full-benchmark score well enough to speed up iteration and
to compare candidates on the Pareto frontier of quality vs. cost/latency?
Operationally, this can be enforced by filtering or weighting configurations by
relevance (model family, recency, tool availability), and our scripts already
support configuration-family filtering (prefix) and explicit exclusions.

\section{Ridge regression and conditioning}
\label{app:ridge}
With a small number of configurations and many candidate tasks ($p \gg n$),
ordinary least squares (OLS) is ill-conditioned: many tasks are redundant or
strongly correlated, so small changes in the observed outcomes (or the
particular train/test split) can produce large swings in fitted coefficients.
Ridge regression addresses this by adding an $\ell_2$ regularization term that
\emph{shrinks} coefficients toward zero.
Intuitively, Ridge prefers explanations of $y$ that use \emph{moderate} weights
spread across correlated tasks rather than extremely large positive/negative
weights that cancel each other.
This improves stability (and often generalization) by reducing estimator
variance, at the cost of introducing a small bias (a standard bias--variance
tradeoff).
In our setting, these properties matter for two practical reasons.
First, the design matrix $\tilde{X}$ of (largely binary) task outcomes can make
$\tilde{X}^\top\tilde{X}$ nearly singular when $p>n$ or when many tasks are
near-duplicates; Ridge yields a unique, well-behaved solution in this regime.
Second, we interpret the resulting coefficients as an
\emph{importance weighting} over the selected tasks, producing a weighted proxy
score that is less sensitive to noise than an unregularized fit.

\paragraph{Objective.}
Given a candidate set of columns $S$, we fit
\begin{equation}
 \hat{\beta}(S) = \arg\min_{\beta}\; \|y - \tilde{X}_S\beta\|_2^2 + \alpha\|\beta\|_2^2,
\end{equation}
where $\tilde{X}_S$ denotes the imputed matrix restricted to columns in $S$ and
$\alpha>0$ is a regularization constant.
The penalty $\alpha\|\beta\|_2^2$ yields the closed-form solution
$\hat{\beta}=(\tilde{X}_S^\top\tilde{X}_S+\alpha I)^{-1}\tilde{X}_S^\top y$,
which is well-defined even when $\tilde{X}_S^\top\tilde{X}_S$ is singular.
Increasing $\alpha$ makes the fit less sensitive to idiosyncratic patterns in
small datasets (more shrinkage), while decreasing it approaches OLS (less
shrinkage, higher variance); we use $\alpha=0.01$ throughout.

\section{Greedy forward selection procedure}
\label{app:greedy}
Selecting the best subset of $k$ tasks is combinatorial, so we build $S$ with
greedy forward selection.
Starting from $S_0=\emptyset$, at each step $t$ we evaluate every remaining
candidate task $q$ by fitting Ridge on $S_{t-1}\cup\{q\}$ under 5-fold inner
cross-validation (shuffled $K$-fold, fixed seed) and computing the
cross-validated (CV) $R^2$. The candidate that yields the highest inner CV $R^2$
is added to $S_t$:
\begin{equation}
 q^* = \arg\max_{q\notin S_{t-1}} \; R^2_{\mathrm{innerCV}}(S_{t-1}\cup\{q\}).
\end{equation}
We stop once $|S|=k$, or when the inner CV $R^2$ exceeds a convergence
threshold.
Besides being tractable, the procedure produces a ranked list that practitioners
can inspect.
The diversity tie-breaker of Section~\ref{sec:test-selection} modifies only the
selection rule $q^*$, replacing the single best candidate with a
diversity-maximizing one drawn from the $\epsilon$-band of near-best candidates.

\begin{algorithm}[!ht]
\SetKwInput{KwInput}{Input}                
\SetKwInput{KwOutput}{Output}              
\DontPrintSemicolon
  \KwInput{Task Set Q}
  \KwOutput{Greedy Selected Task Set $S_k$}

  $S_0 = \phi$ \;
  \For{t in $\{1...k\}$} {
    
    $q_t^* = \arg\max_{q\in Q\setminus S_{t-1}}  R^2_{\mathrm{innerCV}}(S_{t-1}\cup\{q\})$\;
    $S_t = S_{t-1} \cup \{q_t^*\}$\;
 
}
\KwRet $S_k$

\caption{Greedy Selection}
\end{algorithm}

\begin{algorithm}[!ht]
\SetKwInput{KwInput}{Input}                
\SetKwInput{KwOutput}{Output}              
\DontPrintSemicolon
  \KwInput{Task Set $Q$, Tolerance Band $\epsilon$}
  \KwOutput{Selected Task Set $S_k$}

  $S_0 = \phi$ \;
  \For{t in $\{1...k\}$} {
    $R^2_{\mathrm{best}} = \max_{q\in Q\setminus S_{t-1}}R^2_{\mathrm{innerCV}}{(S_{t-1}\cup\{q\})}$ \;
    $ B_\epsilon(S_{t-1}) = \{q\in Q \setminus S_{t-1} : R^2_{\mathrm{innerCV}}(S_{t-1}\cup\{q\}) \ge R^2_{\mathrm{best}} - \epsilon\}$\;
    $q_t^* = \arg\max_{q\in B_\epsilon(S_{t-1})}  \mathrm{Div}(S_{t-1}\cup\{q\})$\;
    $S_t = S_{t-1} \cup \{q_t^*\}$\;
 
}
\KwRet $S_k$

\caption{Modified Greedy Selection}
\end{algorithm}

\section{Taxonomy diversity score $\mathrm{Div}(S)$}
\label{app:diversity}
Each taxonomy dimension $j\in\mathcal{D}$ contributes a coverage score
$\mathrm{Div}_j(S)\in[0,1]$ to the diversity score $\mathrm{Div}(S)$ of
Section~\ref{sec:test-selection}, with a form that depends on its measure type:
\begin{equation}
\mathrm{Div}_j(S) =
\begin{cases}
r_j(S)\,/\,r_j(\mathcal{T}) & \text{(numeric)}, \\[0.5ex]
H(P_S)\,/\,\log k_{\mathcal{T}} & \text{(categorical)}, \\[0.5ex]
|L_j(S)|\,/\,|L_j(\mathcal{T})| & \text{(multi-label)},
\end{cases}
\end{equation}
where, over a set of tasks $A$,
$r_j(A)=\max_{t\in A}x_{t,j}-\min_{t\in A}x_{t,j}$ is the spread of $j$'s
numeric values and $L_j(A)=\bigcup_{t\in A}L_{t,j}$ the set of labels present;
$P_S$ is the selected tasks' distribution over $j$'s categories and
$k_{\mathcal{T}}$ the number of those categories globally.
Numeric measures (e.g., table and workbook size, clarity, task difficulty) use
range coverage; single-label categorical measures (e.g., header type, table
structure, task type, domain) use entropy; and multi-label measures (e.g., meta
operations and required LLM capabilities) use label coverage.
Where a dimension is captured by several numeric signals (e.g., workbook size
and structural complexity), we summarize them by their top principal components,
which we use as numeric measures.

\section{Outer-split validation strategies}
\label{app:splits}
The three outer-loop strategies of Section~\ref{sec:eval-setup} differ in how
configurations are partitioned:
\begin{itemize}[leftmargin=*]
  \item \textbf{Nested $K$-Fold.}
  Configurations are shuffled and partitioned into 5 outer folds.
  In each outer fold the training configurations are used exclusively for
  selection and fitting, the held-out configurations only for prediction, and
  the final $R^2$ is computed from the concatenated out-of-fold predictions.
  \item \textbf{Walk-Forward Expanding.}
  Configurations are sorted chronologically by build id and partitioned into
  equal-size sequential windows.
  The outer loop advances forward in time: fold $i$ trains on windows
  $1,\dots,i$ and tests on window $i+1$.
  \item \textbf{Walk-Forward Sliding.}
  Identical to expanding, but the training window has a fixed size equal to
  twice the test window, so that if the test window covers indices $[t, t+w)$,
  the training window covers $[\max(0, t-2w), t)$.
  Older data is dropped as evaluation advances, so each fold trains on a limited
  recent history.
\end{itemize}

\section{On mean-imputation}
\label{app:imputation}
We impute each missing entry with its per-task mean
$\mu_t = \frac{1}{|\mathcal{C}_t|}\sum_{c\in\mathcal{C}_t} X_{c,t}$ (the task's
average outcome over the configurations $\mathcal{C}_t$ that ran it), writing
$\tilde{X}$ for the resulting dense matrix ($\tilde{X}_{c,t}=X_{c,t}$ when
observed and $\mu_t$ otherwise), so that configurations can be compared on common ground.
Within each outer validation fold we compute $\mu_t$ over the training
configurations only and identify zero-variance tasks on the training fold
alone, so a held-out configuration never contributes to the values imputed for
it.
The final selection on all data uses means over all configurations.
The RQ4 broader analysis (Section~\ref{sec:results-controls}) follows the same
training-fold convention.
Configurations retained by the \texttt{min\_completion\_rate} filter are still
mean-imputed at the task level in exactly this way.
The obvious alternatives to mean-imputation are each biased in this log regime:
dropping configurations with any missing outcome discards most data when $p$ is
large; dropping ever-missing tasks favors those that happen to run more often;
and averaging over only the observed selected tasks introduces a
\emph{variable denominator}, scoring different configurations on different
effective subsets.
A learned imputer (e.g., $k$-nearest-neighbors or low-rank matrix completion)
would instead add a genuine second model that stacks prediction errors into the
regression's inputs and leaks information across tasks.
Mean-imputation is the simplest well-defined baseline.
It is, moreover, a placeholder rather than a predictive step: $\mu_t$ ignores
everything about configuration $c$ and is identical for every configuration
missing task $t$, so an imputed cell cannot help separate configurations---it
only dilutes the signal in the observed entries, never manufactures new signal.
The sole predictive step is the Ridge fit over the selected columns; the cost of
imputation is statistical rather than architectural, attenuating the fitted
coefficients of heavily-missing columns (a bias we revisit in
Section~\ref{sec:discussion}).

\section{Matched-budget controls on the complete-case matrix}
\label{app:complete-case-controls}
Table~\ref{tab:complete-case-baselines} compares taxonomy-guided selection
with the five controls of Section~\ref{sec:sweep-grid} at $k{=}10$ on the
zero-missing common-reference matrix (99 execution IDs by 715 common tasks,
595 nonconstant candidates); Section~\ref{sec:results-controls} summarizes the
result.
At $\epsilon{=}0.05$, taxonomy-guided selection has the highest honest $R^2$
under all three validation strategies, and at both $\epsilon$ values it exceeds
its matched $\epsilon$-band-random control under all three.
At $\epsilon{=}0.10$ its diversity matches that of diversity-only selection,
which has the lowest honest $R^2$ of any method.

\begin{table*}[t]
\centering
\caption{Matched-budget comparison on the zero-missing common-reference
rectangle (99 execution IDs, 715 common tasks, 595 nonconstant candidates,
$k=10$). Values are honest weighted outer-validation $R^2$. Randomized
methods report the mean over three seeds. Diversity is measured on the suite
selected using the full complete-case dataset and is averaged across seeds
for randomized methods.}
\label{tab:complete-case-baselines}
\small
\begin{tabular}{lccccc}
\toprule
Method & $\epsilon$ & Nested $R^2$ & Expanding $R^2$
& Sliding $R^2$ & Diversity \\
\midrule
Greedy predictive       & --   & 0.689 & 0.379 & 0.472 & 0.649 \\
Taxonomy-guided         & 0.10 & 0.693 & 0.482 & 0.479 & \textbf{0.820} \\
Taxonomy-guided         & 0.05 & \textbf{0.731} & \textbf{0.493}
& \textbf{0.496} & 0.815 \\
$\epsilon$-band random  & 0.10 & 0.569 & 0.277 & 0.225 & 0.485 \\
$\epsilon$-band random  & 0.05 & 0.626 & 0.275 & 0.324 & 0.588 \\
Uniform random          & --   & 0.567 & 0.301 & 0.338 & 0.490 \\
Diversity-only          & --   & 0.490 & -0.014 & 0.067 & \textbf{0.820} \\
Taxonomy-stratified     & --   & 0.602 & 0.205 & 0.217 & 0.665 \\
\bottomrule
\end{tabular}
\end{table*}

\clearpage
\section{Honest $R^2$ over the full $k \times$ \texttt{min\_completion\_rate} grid}

Figures~\ref{fig:kmin-grid-kfold}--\ref{fig:kmin-grid-wfsliding2} report the
honest weighted-Ridge $R^2$ vs.\ $\epsilon$ across the full $k \times$
\texttt{min\_completion\_rate} grid, one figure per validation strategy.
Across all three, the regularization benefit of moderate $\epsilon$ is most
pronounced at low \texttt{min\_completion\_rate} and small $k$.

\begin{figure*}[p]
\centering
\includegraphics[height=\textheight]{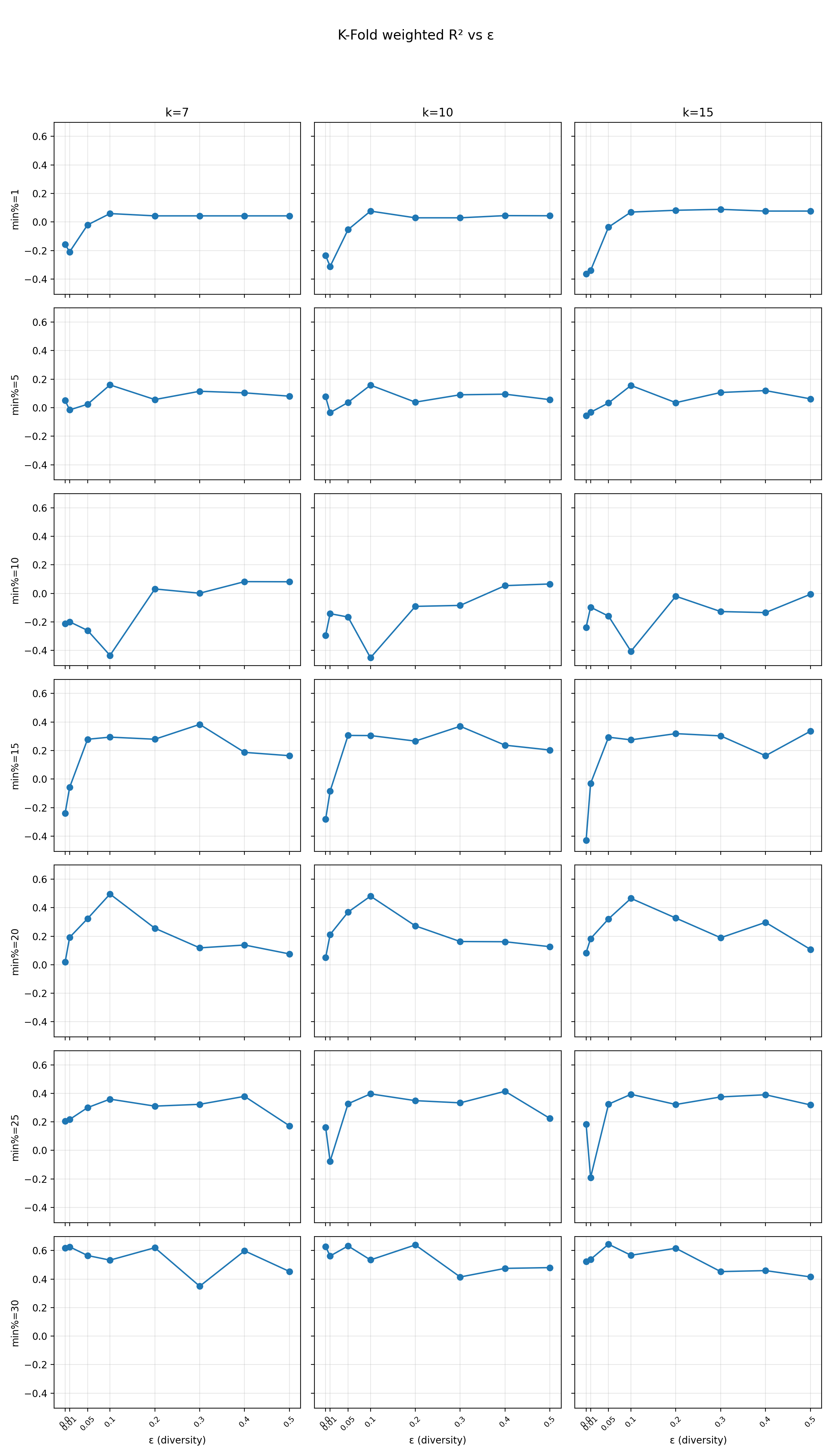}
\caption{Nested $K$-Fold: honest weighted-Ridge $R^2$ vs.\ $\epsilon$ across $k \times$ \texttt{min\_completion\_rate} grid for $k \in \{7, 10, 15\}$.}
\label{fig:kmin-grid-kfold}
\end{figure*}

\begin{figure*}[p]
\centering
\includegraphics[height=\textheight]{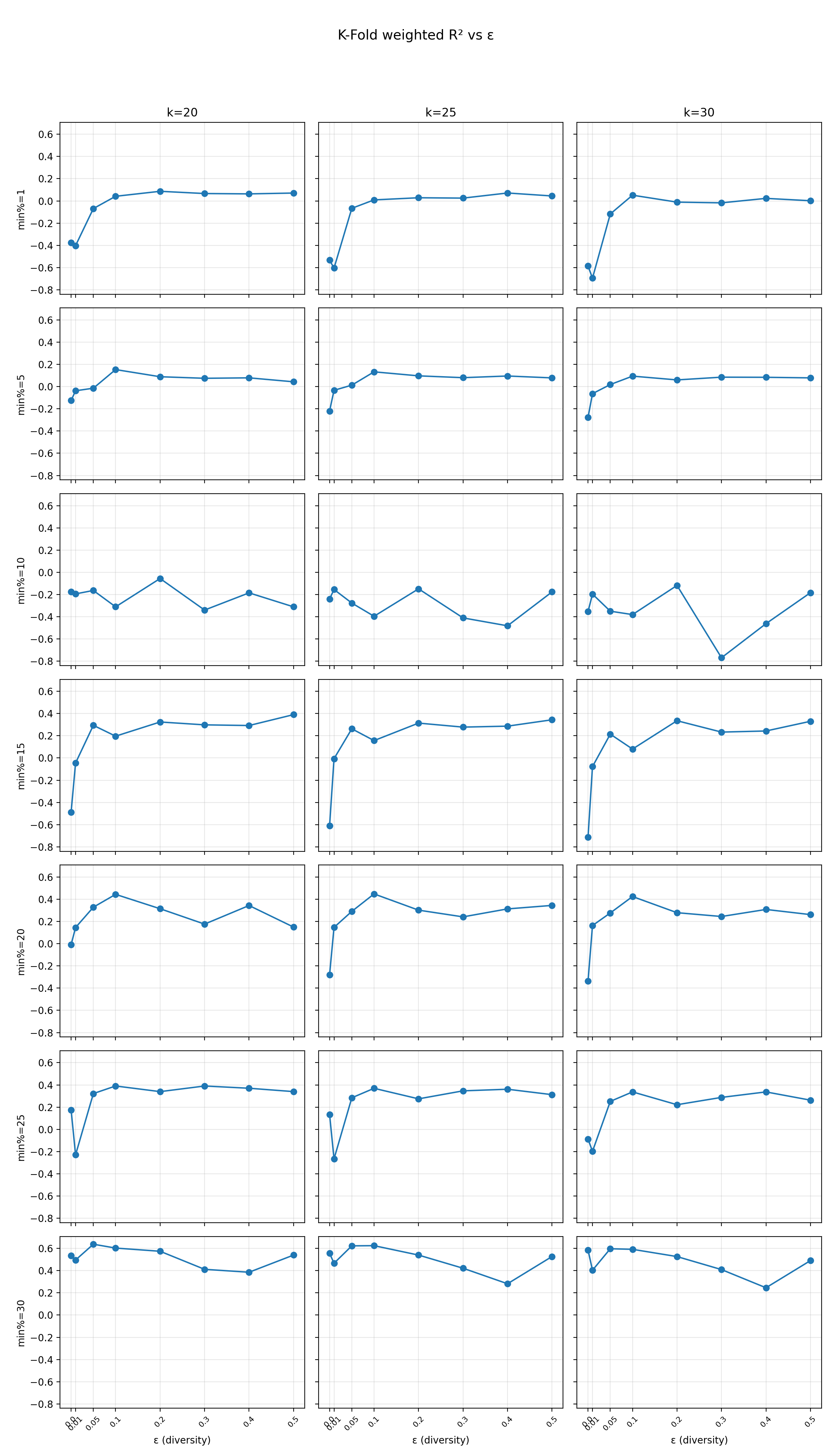}
\caption{Nested $K$-Fold: honest weighted-Ridge $R^2$ vs.\ $\epsilon$ across $k \times$ \texttt{min\_completion\_rate} grid for $k \in \{20, 25, 30\}$.}
\label{fig:kmin-grid-kfold2}
\end{figure*}

\begin{figure*}[p]
\centering
\includegraphics[height=\textheight]{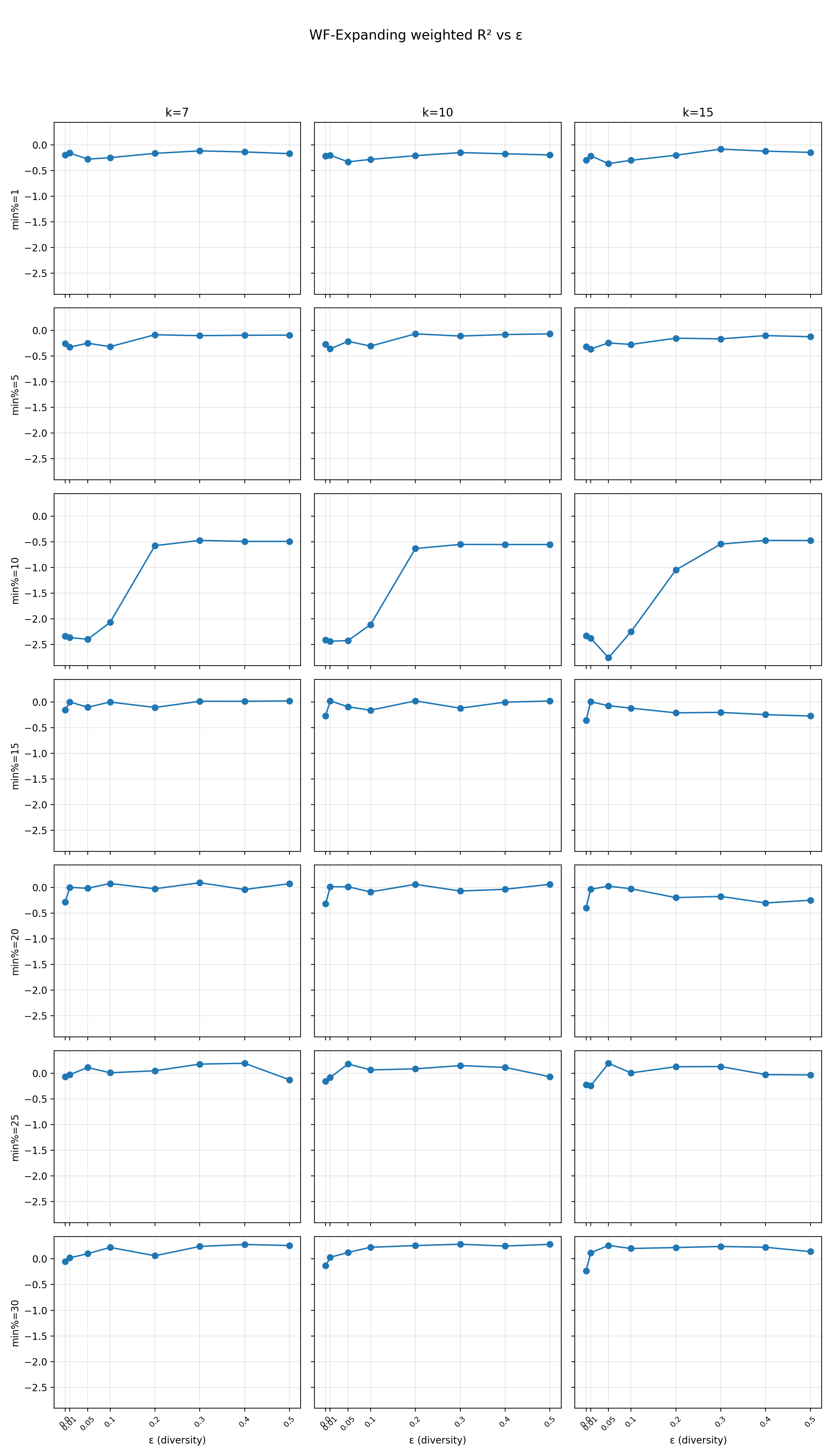}
\caption{Walk-Forward Expanding: honest weighted-Ridge $R^2$ vs.\ $\epsilon$ across $k \times$ \texttt{min\_completion\_rate} grid for $k \in \{7, 10, 15\}$.}
\label{fig:kmin-grid-wfexpanding}
\end{figure*}

\begin{figure*}[p]
\centering
\includegraphics[height=\textheight]{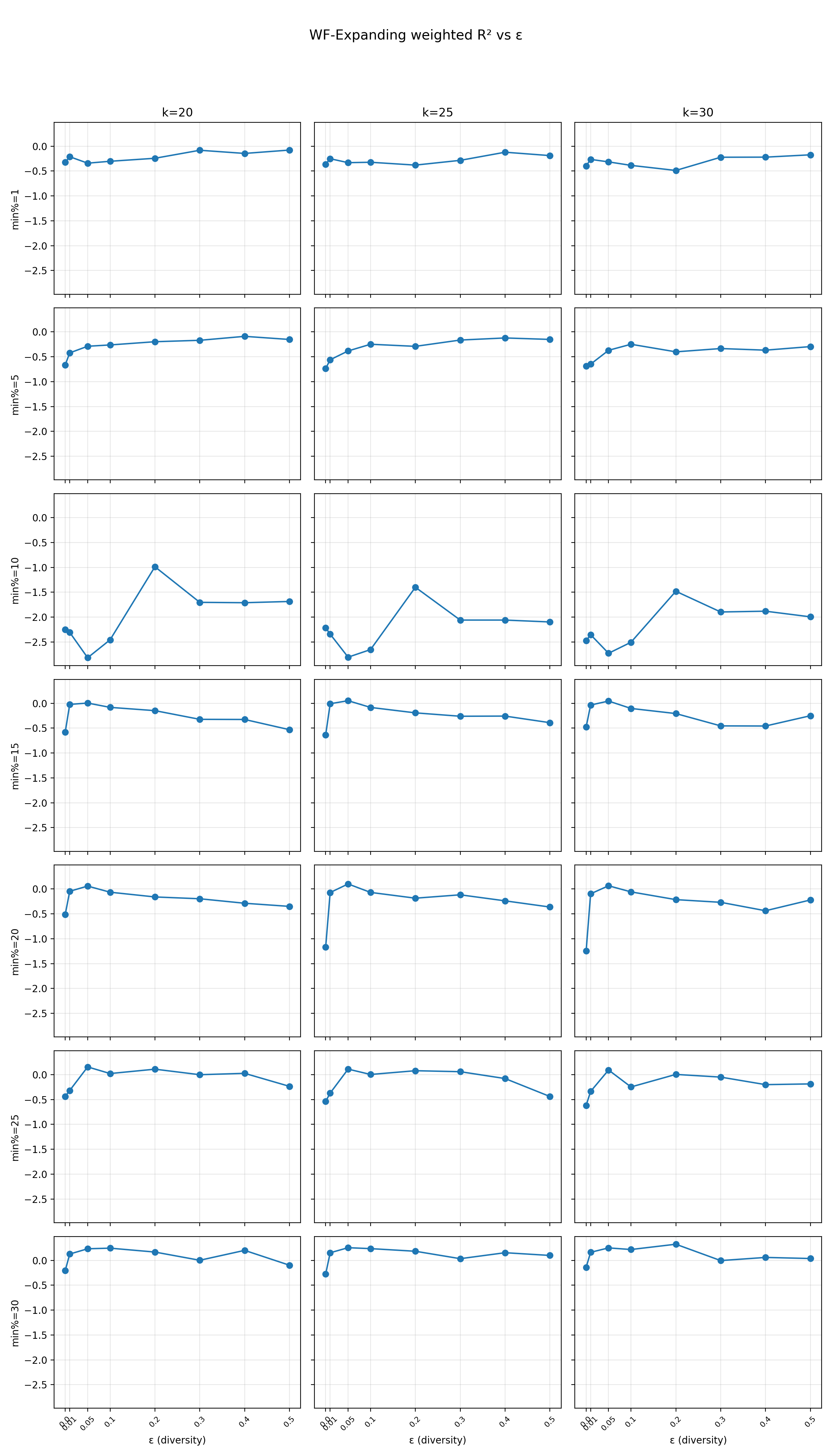}
\caption{Walk-Forward Expanding: honest weighted-Ridge $R^2$ vs.\ $\epsilon$ across $k \times$ \texttt{min\_completion\_rate} grid for $k \in \{20, 25, 30\}$.}
\label{fig:kmin-grid-wfexpanding2}
\end{figure*}

\begin{figure*}[p]
\centering
\includegraphics[height=\textheight]{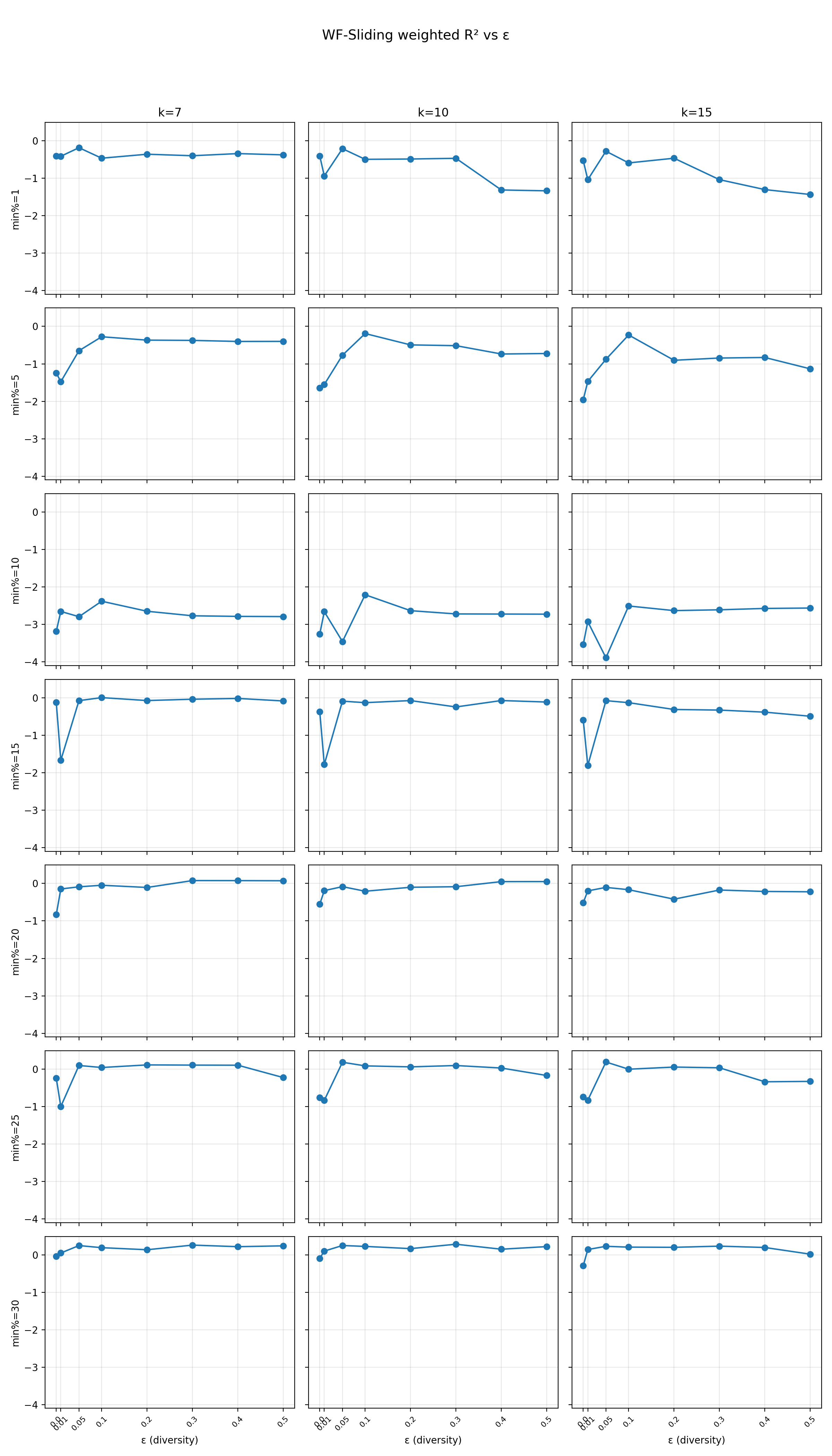}
\caption{Walk-Forward Sliding: honest weighted-Ridge $R^2$ vs.\ $\epsilon$ across $k \times$ \texttt{min\_completion\_rate} grid for $k \in \{7, 10, 15\}$.}
\label{fig:kmin-grid-wfsliding}
\end{figure*}

\begin{figure*}[p]
\centering
\includegraphics[height=\textheight]{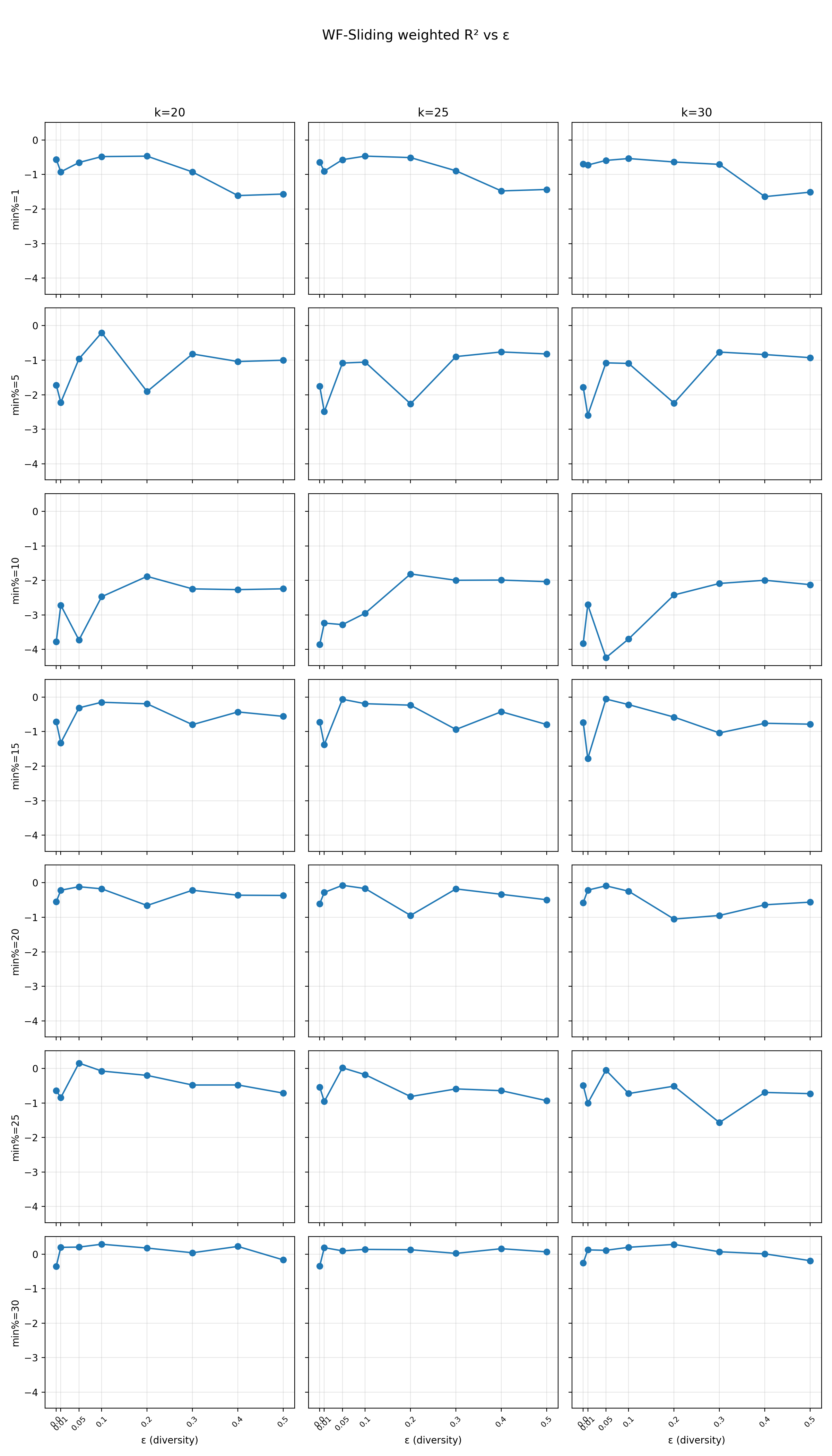}
\caption{Walk-Forward Sliding: honest weighted-Ridge $R^2$ vs.\ $\epsilon$ across $k \times$ \texttt{min\_completion\_rate} grid for $k \in \{20, 25, 30\}$.}
\label{fig:kmin-grid-wfsliding2}
\end{figure*}

\end{document}